\pdfoutput=1

\documentclass[10pt,twoside]{article}
\usepackage[super,sort&compress,comma]{natbib} 
\usepackage[version=3]{mhchem} 
\usepackage{times,mathptm}
\usepackage{sectsty}
\usepackage{balance} 
\usepackage[textwidth=16.5cm,textheight=24.4cm,centering]{geometry}
\usepackage{amsmath}

\usepackage{graphicx} 
\usepackage{float} 
\usepackage[format=plain,justification=justified,singlelinecheck=false,font={stretch=1.125,small,sf},labelfont=bf,labelsep=space,belowskip=-10pt]{caption}
\usepackage{setspace}
\usepackage{droidsans}
\usepackage{charter}
\usepackage[T1]{fontenc}
\usepackage{extsizes}
\usepackage{fancyhdr}
\usepackage{multirow} 
\usepackage{stackengine} 
\usepackage{makecell} 
\usepackage{url} 

\usepackage{parskip}
\usepackage{mathtools} 
\usepackage{dblfloatfix} 

\usepackage{siunitx}
\DeclareSIUnit\angstrom{\text {Å}}

\usepackage{graphicx}
\usepackage{dcolumn}
\usepackage{bm}
\usepackage[capitalise]{cleveref} 
\usepackage{natbib} 
\usepackage{xfrac} 

\usepackage{xargs}                      
\usepackage[pdftex,dvipsnames]{xcolor}  
\usepackage[colorinlistoftodos,prependcaption,textsize=tiny]{todonotes}
\newcommandx{\unsure}[2][1=]{\todo[linecolor=red,backgroundcolor=red!25,bordercolor=red,#1]{#2}}
\newcommandx{\change}[2][1=]{\todo[linecolor=blue,backgroundcolor=blue!25,bordercolor=blue,#1]{#2}}
\newcommandx{\info}[2][1=]{\todo[linecolor=OliveGreen,backgroundcolor=OliveGreen!25,bordercolor=OliveGreen,#1]{#2}}
\newcommandx{\improvement}[2][1=]{\todo[linecolor=Plum,backgroundcolor=Plum!25,bordercolor=Plum,#1]{#2}}
\newcommandx{\thiswillnotshow}[2][1=]{\todo[disable,#1]{#2}}

\newcommand{\super}[1]{$^{\text{#1}}$}
\newcommand{\sub}[1]{$_{\text{#1}}$}
\newcommand{\image}[3][]{%
  \begin{figure}[H]%
    \centering%
    \includegraphics[height={#3}]{{#2}.eps}%
    \ifx&#1&
    \else
      \caption*{#1}
    \fi
  \end{figure}%
}
\newcommand{\img}[2][]{
  \begin{figure}[H] 
    \centering 
    \includegraphics{{#2}.eps} 
    \ifx&#1&
    \else
      \caption*{#1}
    \fi
  \end{figure}
}

\usepackage{pdftexcmds}
\makeatletter
\newcommand\kv[2]{%
  \ifnum\pdf@strcmp{\unexpanded{#1}}{V}=0 %
     \expandafter\@firstoftwo
  \else
    \expandafter\@secondoftwo
  \fi
    {\textit{#1}\sub{#2}}
    {#1\sub{#2}}%
}
\makeatother
\makeatletter
\newcommand\kvc[3]{%
  \ifnum\pdf@strcmp{\unexpanded{#1}}{V}=0 %
     \expandafter\@firstoftwo
  \else
    \expandafter\@secondoftwo
  \fi
    {\textit{#1}\sub{#2}\super{#3}\negmedspace}
    {#1\sub{#2}\super{#3}\negmedspace}
}
\makeatother

\DeclareMathVersion{mathchartertext}
\SetSymbolFont{letters}{mathchartertext}{OML}{mdbch}{m}{n}

\makeatletter
\newcommand*{\centerfloat}{
  \parindent \z@
  \leftskip \z@ \@plus 1fil \@minus \textwidth
  \rightskip\leftskip
  \parfillskip \z@skip}
\makeatother

\makeatletter
\DeclareRobustCommand\citenum
   {\begingroup
     \NAT@swatrue\let\NAT@ctype\z@\NAT@parfalse\let\textsuperscript\relax
     \NAT@citexnum[][]}
\makeatother

\begin{document}

\thispagestyle{plain}
\fancypagestyle{plain}{
\renewcommand{\headrulewidth}{1pt}}
\renewcommand{\thefootnote}{\fnsymbol{footnote}}
\renewcommand\footnoterule{\vspace*{1pt}%
\hrule width 15cm height 0.4pt \vspace*{5pt}} 
\setcounter{secnumdepth}{5}

\makeatletter 
\renewcommand{\fnum@figure}{\textbf{Fig.~\thefigure~~}}
\def\subsubsection{\@startsection{subsubsection}{3}{10pt}{-1.25ex plus -1ex minus -.1ex}{0ex plus 0ex}{\normalsize\bf}} 
\def\paragraph{\@startsection{paragraph}{4}{10pt}{-1.25ex plus -1ex minus -.1ex}{0ex plus 0ex}{\normalsize\textit}} 
\renewcommand\@biblabel[1]{#1}            
\renewcommand\@makefntext[1]%
{\noindent\makebox[0pt][r]{\@thefnmark\,}#1}
\makeatother 
\sectionfont{\sffamily\large}
\subsectionfont{\normalsize} 
\setstretch{1.125} 
\makeatletter
\renewcommand\LARGE{\@setfontsize\LARGE{15pt}{17}}
\renewcommand\Large{\@setfontsize\Large{12pt}{14}}
\renewcommand\large{\@setfontsize\large{10pt}{12}}
\renewcommand\footnotesize{\@setfontsize\footnotesize{7pt}{10}}
\renewcommand\scriptsize{\@setfontsize\scriptsize{7pt}{7}}
\makeatother
\renewcommand{\figurename}{\small{Fig.}~}
\sectionfont{\Large}
\subsectionfont{\normalsize}
\subsubsectionfont{\bf}

\fancyfoot{}
\fancyfoot[RO]{\scriptsize{\sffamily{1--12 ~\textbar  \hspace{2pt}\thepage}}} 
\fancyfoot[LE]{\scriptsize{\sffamily{\thepage~\textbar 1--12}}}
\fancyhead{}
\newgeometry{textwidth=0in}
\fancyheadoffset{\textwidth}
\renewcommand{\headrulewidth}{1pt} 
\renewcommand{\footrulewidth}{1pt}
\setlength{\arrayrulewidth}{1pt}
\setlength{\columnsep}{6.5mm}
\setlength\bibsep{1pt}
\newgeometry{textwidth=16.5cm,textheight=24.4cm,centering,headsep=10pt,footskip=24pt}

\noindent\LARGE{\textbf{\vspace{-0.2cm}\\Impact of metastable defect structures on carrier recombination in solar cells$^\dag$}}
\vspace{0.3cm}

\noindent\large{\textbf{Seán R. Kavanagh,$^{\ast}$\textit{$^{a,b}$} David O. Scanlon,\textit{$^{a}$} Aron Walsh\textit{$^{a,c}$} and
Christoph Freysoldt\textit{$^{d}$}}}\vspace{0.5cm}


\renewcommand*\rmdefault{bch}\normalfont\upshape
\rmfamily
\section*{}
\vspace{-2cm}

\footnotetext{\textit{$^{a}$~Department of Chemistry \& Thomas Young Centre, University College London, 20 Gordon Street, London WC1H 0AJ, UK}}
\footnotetext{\textit{$^{b}$~Department of Materials \& Thomas Young Centre, Imperial College London, Exhibition Road, London SW7 2AZ, UK}}
\footnotetext{\textit{$^{c}$~Department of Materials Science and Engineering, Yonsei University, Seoul 03722, Republic of Korea}}
\footnotetext{\textit{$^{d}$~Max-Planck-Institut für Eisenforschung GmbH, Max-Planck-Str. 1, 40237 Düsseldorf, Germany}}
\footnotetext{\textit{$^{\ast}$~sean.kavanagh.19@ucl.ac.uk}}

\footnotetext{\dag~Electronic Supplementary Information (ESI) appended.}


\vspace{0.3cm}
\noindent \normalsize{
The efficiency of a solar cell is often limited by electron-hole recombination mediated by defect states within the band gap of the photovoltaic (PV) semiconductor. The Shockley-Read-Hall (SRH) model considers a static trap that can successively capture electrons and holes. In reality however, true trap levels vary with both the defect charge state and local structure. Here we consider the role of metastable structural configurations in capturing electrons and holes, taking the tellurium interstitial in CdTe as an illustrative example. Consideration of the defect dynamics, and symmetry-breaking, changes the qualitative behaviour and activates new pathways for carrier capture. Our results reveal the potential importance of metastable defect structures in non-radiative recombination, in particular for semiconductors with anharmonic/ionic-covalent bonding, multinary compositions, low crystal symmetries or highly-mobile defects.
}
\setcounter{secnumdepth}{0} 
\rmfamily

\textit{Submitted to Faraday Discussions, to be presented at the RSC Faraday Discussion on 'Emerging inorganic materials in thin-film photovoltaics' in Bath, UK, July 2022.}


\section{Introduction}
The search for new thin-film photovoltaic (PV) materials must balance advantageous properties (such as high absorbance, a band gap matching the targeted light spectrum, easy separation of electrons and holes, etc.) with the susceptibility to performance-limiting loss mechanisms.
Among these, defect-mediated non-radiative recombination represents the dominant loss mechanism in emerging inorganic PV technologies.\cite{huang_perovskite-inspired_2021,rau_efficiency_2017}
Identifying the `killer' defect traps and hence potential synthesis and processing strategies to avoid their effects is crucial to mitigating losses and achieving high performance.
For instance, the major efficiency improvements of the last 5 years for single-junction perovskite solar cells can be directly attributed to effective trap management.\cite{jin_its_2020}

Point defects, such as vacancies, interstitials and antisites, facilitate carrier recombination by breaking the periodicity of the solid and thus introducing new electronic states within the semiconducting band gap.
By successively capturing electrons and holes through non-radiative (vibrational) processes, these defect levels allow excited charge carriers to recombine
across the gap, reducing the open-circuit voltage in solar cells and the quantum efficiencies of light-emitting diodes for example.\cite{huang_perovskite-inspired_2021}
Recent advances in both theory and computation have rendered possible the explicit prediction of defect recombination activity,\cite{alkauskas_first-principles_2014,sunghyun_kim_carriercapturejl_2020,turiansky_nonrad_2021,kim_anharmonic_2019,shi_ab_2012,kavanagh_rapid_2021} allowing for the \textit{ab initio} identification of such `killer' traps.
The synergistic combination of theoretical methods and experimental characterisation has allowed for rapid improvements in photoconversion efficiencies for emerging inorganic PV materials.\cite{zakutayev_emerging_2021,huang_perovskite-inspired_2021}

Point defects can exhibit complex potential energy surfaces (PESs), with multiple local minima.\cite{mosquera-lois_search_2021,kavanagh_rapid_2021,krajewska_enhanced_2021,goes_identification_2018, lany_metal-dimer_2004}
As such, their atomic structures often do not reside solely in their global equilibrium configuration, but may access alternative metastable structures via thermal, photo or electronic conversion.
The possibility of structural transformation at defect sites has several potential impacts for technological applications --- for example, low energy barriers between structures could lead to coherence corruption and a breakdown in performance for defects used in quantum computing and single-photon emission.
In particular, carrier capture rates at defects are sensitively dependent on the atomic structure, varying by over 15 orders of magnitude with different geometries for the same nominal defect species,\cite{kavanagh_rapid_2021,alkauskas_role_2016} and so the presence of these alternative configurations can introduce alternative pathways which significantly alter the overall recombination kinetics. 
Indeed, this phenomenon has recently been reported in the literature, with metastable defects accelerating the electron-hole recombination process and transforming benign defect centres into harmful traps.\cite{kavanagh_rapid_2021,alkauskas_role_2016,guo_nonradiative_2021,yang_non-radiative_2016}

A complete analysis of the possible mechanisms and resulting impact of metastable configurations on non-radiative recombination in semiconductors is lacking.
In this work, we first provide a general analysis of the potential recombination pathways introduced by metastable defect structures.
We discuss the conditions for this behaviour to occur, the anticipated trends and the factors governing the consequences for material properties.
We highlight tellurium interstitials in CdTe as illustrative exemplars of the potential complexity and importance of metastable configurations to carrier recombination, exhibiting a novel thermal-excitation recombination cycle which has not previously been reported in the literature, to the knowledge of the authors.
Finally, we discuss the important implications of these findings to the broader field of photovoltaic materials research.


\section{Metastable Defect Dynamics}
\begin{figure}[h]
\centering
\includegraphics[width=0.9\textwidth]{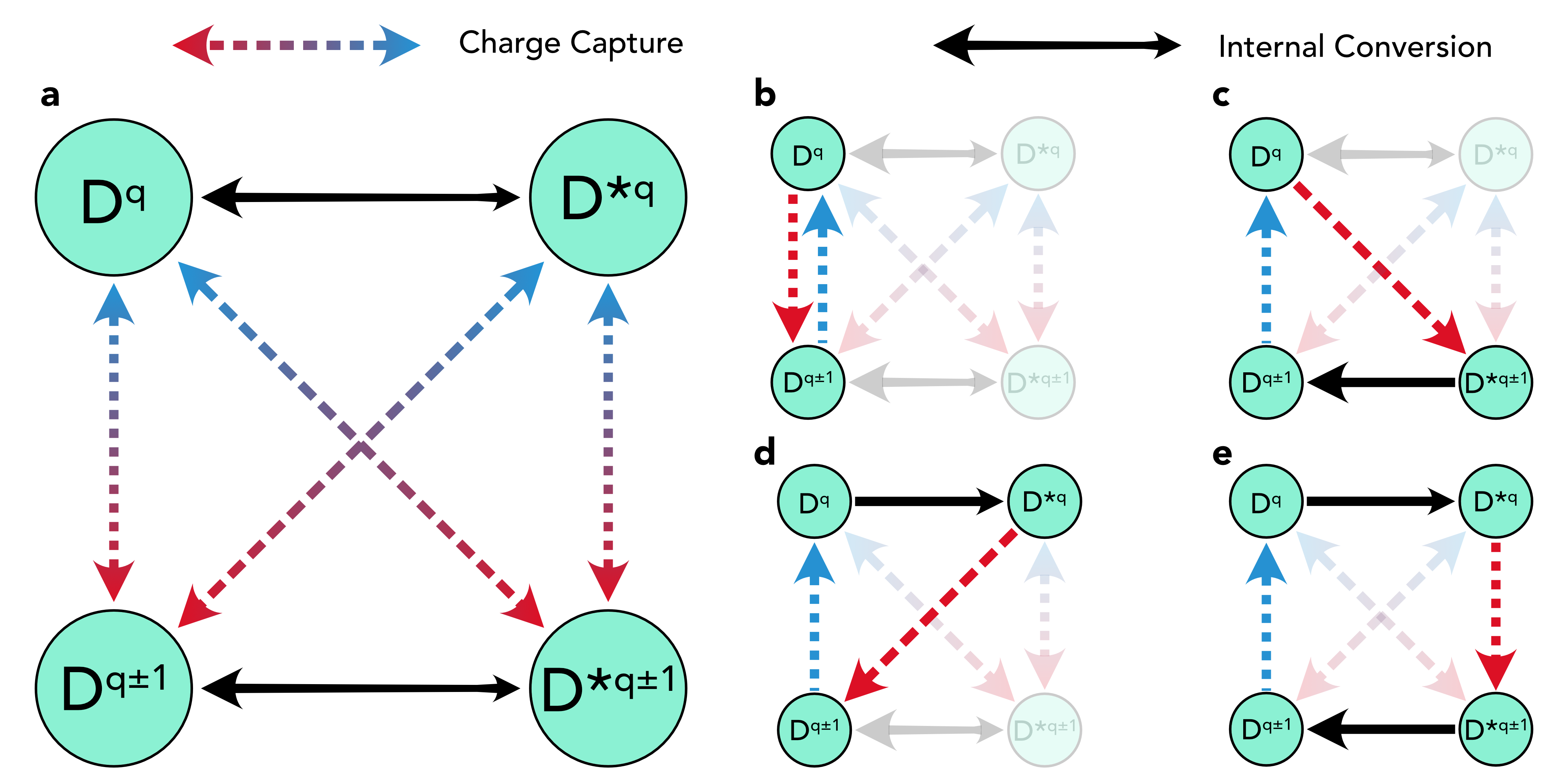}
\caption{\textbf{a.} Graph network illustration of the potential structural transitions involving charge capture (red/blue) or internal conversion (black) for a defect $D$ in charge state $q$. A single metastable configuration $D^{*\,q}$ is assumed to be accessible for each charge state.
\textbf{b.} Pathway for a typical 2-step Shockley-Read-Hall recombination process.
\textbf{c.} Pathway for a 3-step recombination process involving \emph{capture into} a metastable defect structure, followed by internal relaxation and charge capture back to the initial state.
\textbf{d.} Pathway for a 3-step recombination process involving \emph{internal excitation} to a metastable structure, followed by electron and hole charge capture.
\textbf{e.} Pathway for a 4-step recombination process involving \emph{internal excitation} to a metastable structure ($D^{*\,q}$), charge capture to the $q\pm1$ metastable structure ($D^{*\,q\pm1}$), \emph{internal relaxation} and charge capture back to the initial configuration ($D^{\,q}$)}
\label{fig:potential_transitions}
\end{figure}

There are several mechanisms by which the presence of metastable structures can influence the overall defect behaviour in solids.
Regarding non-radiative carrier recombination, the structural transition pathways introduced by metastable configurations on the defect potential energy surface can act to accelerate or retard this process, or may not affect it at all.
\cref{fig:potential_transitions} illustrates these possible transitions using a simplified graph network for the case of a single metastable configuration $D^*$ that is both accessible and significantly influential on the overall capture behaviour.

The impact of these metastable defects depends 
on the competing charge capture rates (dashed red/blue arrows),
the relative energies \emph{of} and transition barriers \emph{between} ground and metastable configurations (i.e. their accessibility)(solid black arrows) and the defect energy level position in the host band gap (\cref{eq:charge_transition_level}).
Each of the charge capture transitions can be fast or slow, with no \emph{necessary} correlation between opposing directions.
That said, the recurrence of harmonic PESs means that often a fast electron capture rate implies slow hole capture in the reverse direction, and vice versa.\cite{huang_perovskite-inspired_2021,das_what_2020,kavanagh_rapid_2021} 
This is typically the result of the defect level lying close to the valence or conduction band, often yielding strong/weak interaction with the near/distant band edge and thus fast/slow carrier capture.
This is the intuitive rationale behind the expectation that deep midgap defect levels will act as efficient recombination centres.\cite{huang_perovskite-inspired_2021,alkauskas_role_2016,kirchartz_impact_2018,das_what_2020}
In many cases however, this trend is violated as a result of anharmonicity in the defect PES, often arising due to symmetry-breaking in the defect structure.\cite{kavanagh_rapid_2021,kim_anharmonic_2019,zhang_correctly_2020}
The prevalence of this phenomenon in semiconductors is becoming increasingly better understood, owing to advances in experimental characterisation and computational modelling techniques.\cite{huang_perovskite-inspired_2021,mosquera-lois_search_2021,kavanagh_rapid_2021,menendezproupin_tellurium_2015-1,shepidchenko_small_2015,kumagai_insights_2021}
On the other hand, there is a clear and definite correlation between the opposing internal conversion directions, indicated by the relative sizing of the black arrows in \cref{fig:potential_transitions}.
The transition energy barrier from ground to higher energy metastable configuration must be larger than the reverse, hence $k_{\,D \rightarrow D^*} < k_{\,D^* \rightarrow D}$ where $k$ is the transition rate constant (i.e. it is always easier to thermally relax than to excite).

Transitions to higher charge states ($q\pm2$) have not been included here as the behaviour is mostly equivalent and, regardless, for recombination to occur, a closed path must be established (as in \cref{fig:potential_transitions}b-e), returning the defect to the initial configuration.
That said, two ways in which $q\pm2$ defects \emph{could} influence the overall recombination rate here is (1) by introducing rapid capture into, but slow capture out of, these extremal charge states, thus acting to inhibit recombination as defects get `stuck' in inert charge states,\cite{huang_perovskite-inspired_2021,alkauskas_role_2016,zhang_correctly_2020} or (2) the rare scenario where $D^{\,q\pm2}$ introduces an alternative low-barrier pathway between $D^{\,q\pm1}$ and $D^{*\,q\pm1}$ which does not exist otherwise, potentially accelerating the recombination process (i.e. introducing an additional path along the lower rim of the graph network in \cref{fig:potential_transitions}a).

\cref{fig:potential_transitions}c serves as an illustrative example of the potential acceleration or deceleration of the recombination cycle by metastable structures, with the rate of the internal relaxation (black arrow) from $D^{*\,q\pm1}$ to $D^{\,q\pm1}$ being the deciding factor.
Assuming rapid capture transitions for $q \rightarrow q^*\pm1$ (red) and $q\pm1 \rightarrow q$ (blue), a fast internal relaxation means that $D^{*\,q\pm1}$ acts as a reaction intermediate, introducing a rapid pathway for the recombination cycle to proceed.
If, however, internal conversion is extremely slow and $D^{*\,q\pm1}$ does not readily relax to $D^{\,q\pm1}$, several alternative pathways may emerge.
The long-lived metastable configuration could undergo charge capture and return back to the original $D^{\,q}$, completing the \textit{e-h} recombination cycle without involving $D^{\,q\pm1}$.
Such behaviour would increase/decrease the overall recombination rate if the rate-limiting step for $D^{\,q}\Longleftrightarrow D^{*\,q\pm1}$ was faster/slower than that of $D^{\,q}\Longleftrightarrow D^{\,q\pm1}$. 
Instead, if the defect level is sufficiently close to a band edge $D^{*\,q\pm1}$ could readily emit the captured charge, acting as a shallow defect level with no effect on the recombination rate.
Alternatively, if $D^{*\,q\pm1}$ has both slow internal relaxation \emph{and} slow charge capture, defects could become `stuck' in this inert state and in fact greatly diminish the overall recombination activity.

In the above discussion, only non-radiative (phonon emission) pathways are considered.
Under certain conditions, both charge capture and internal conversion may also proceed radiatively through photon emission.
A detailed consideration of such behaviour is beyond the scope of this report, though we note that it is non-radiative capture which dominates efficiency losses in PV semiconductors.\cite{stoneham_theory_2001,huang_perovskite-inspired_2021,alkauskas_first-principles_2014}

%

\subsection{Capture \emph{Into} Metastable Defects}
In recent years, several papers have highlighted the decisive role that metastable defects can play in charge-carrier capture kinetics.\cite{kavanagh_rapid_2021,yang_non-radiative_2016,alkauskas_role_2016,guo_nonradiative_2021}
In each of these studies, it is the situation described in \cref{fig:potential_transitions}c that is reported, where the defect $D^{\,q}$ captures an electron or hole by transforming to a metastable configuration of the $q\pm1$ defect ($D^{*\,q\pm1}$).
This metastable species then relaxes to the ground-state $D^{\,q\pm1}$ through an internal conversion reaction, whether a spin-flip de-excitation\cite{alkauskas_role_2016} or (more commonly) structural/vibrational relaxation,\cite{guo_nonradiative_2021,yang_non-radiative_2016} or both.\cite{kavanagh_rapid_2021}
In each case, this alternative pathway is found to proceed more rapidly than the standard $D^{\,q} \Longleftrightarrow D^{\,q\pm1}$ charge capture process (\cref{fig:potential_transitions}b), thus enhancing the overall recombination rate and reconciling theoretical recombination models with experimental observations.

The question of whether one should expect this to typically be the case, where the presence of a low-lying metastable structure $D^{*\,q\pm1}$ \emph{increases} the recombination rate, is still up for discussion.
The reporting of this phenomenon in the literature is undoubtedly biased, likely only to be discussed when the metastable defect accelerates recombination and ignored otherwise, thus discouraging any general conclusions based on the prevalence of these observations.

\begin{figure}[h]
\centering
\includegraphics[width=0.7\textwidth]{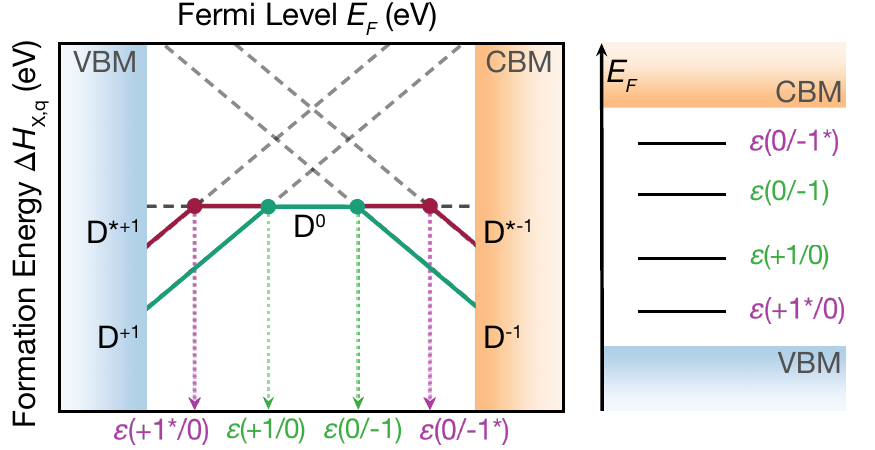}
\caption{Charge transition level positions for ground-state $\varepsilon(q/q\pm1)$ and ground-state $\Longleftrightarrow$ metastable configurations $\varepsilon(q/q^*\pm1)$, on a defect formation energy diagram \textbf{(left)} and a vertical energy level diagram \textbf{(right)}}
\label{fig:metastable_TLs}
\end{figure}

The shift in defect energy levels when metastable defects are involved (depicted in \cref{fig:metastable_TLs}), does however permit some general expectations for the relative capture rates.
As mentioned above, much of the electronic behaviour of point defects is dictated by the position of their energy levels in the semiconducting band gap. 
These in-gap levels are given by their `thermodynamic charge transition level' (TL) positions $\varepsilon(q_1/q_2)$, which are the Fermi level positions where the equilibrium charge state switches between $q_1$ and $q_2$, given by:
\begin{equation}\label{eq:charge_transition_level}
    \varepsilon(q_1/q_2) = \frac{\Delta H_{D,\,q_1}(E_F=0) - \Delta H_{D,\,q_2}(E_F=0)}{q_2 - q_1}
\end{equation}
where $\Delta H_{D,\,q}(E_F=0)$ is the formation energy of defect $D$ with charge $q$ when the Fermi level is at the zero-reference point (the VBM, by convention). 
Taking $q_1 = q$, $q_2 = q\pm1$, and denoting the energy of the metastable defect $\Delta H_{D^*\!\!,\,q\pm1}$ as:
\begin{equation}
    \Delta H_{D^*\!\!,\,q\pm1} = \Delta H_{D,\,q\pm1} + \Delta E 
\end{equation}
where $\Delta E$ is the energy of the metastable defect relative to the ground-state structure; we can then substitute into \cref{eq:charge_transition_level} and rearrange (full derivation in \cref{sisec:Capture_into_meta}$^\dag$) to obtain:
\begin{equation}\label{eq:capture_into_meta_general}
    \varepsilon(q/q^*\pm1) = \varepsilon(q/q\pm1) \mp \Delta E
\end{equation}
\begin{equation}\label{eq:e_capture_into_meta}
    e^-\textrm{ capture} \Longrightarrow\ \ \varepsilon(q/q^*-1) = \varepsilon(q/q-1) + \Delta E
\end{equation}
\begin{equation}\label{eq:h_capture_into_meta}
    h^+\textrm{ capture} \Longrightarrow\ \ \varepsilon(q/q^*+1) = \varepsilon(q/q+1) - \Delta E
\end{equation}
Thus we witness that for charge capture \emph{into} a metastable structure, $D^{\,q} \rightarrow D^{\,*q\pm1}$, the transition level will move an energy $\Delta E$ \emph{closer to} the corresponding band edge (i.e. to higher energy in the band gap for electron capture or to lower energy for hole capture), assuming a transition level $\varepsilon(q/q\pm1)$ initially located within the band gap.
Conversely, this shift in TL position also means 
the TL will be located \emph{further} from the relevant band edge for the $D^{*\,q\pm1} \rightarrow D^{\,q}$ transition (capture \emph{from} the metastable defect).
For example, the $\varepsilon(0/-1^*)$ transition level is located closer to the CBM, from which electrons are captured by $D^{\,0}$ to form $D^{*\,-1}$, than $\varepsilon(0/-1)$, but further from the VBM, from which $D^{*\,-1}$ could capture holes to complete the recombination cycle.
Under the conventional rationale of capture rates being exponentially dependent on the separation of the defect level from the band edge,\cite{huang_perovskite-inspired_2021,das_what_2020,kavanagh_rapid_2021,kirchartz_impact_2018} this implies a faster capture into the metastable configuration ($D^{\,q} \rightarrow D^{*\,q\pm1}$) than the ground-state process ($D^{\,q} \rightarrow D^{\,q\pm1}$), but a slower capture rate in the opposite direction ($D^{*\,q\pm1} \rightarrow D^{\,q}$).
As discussed above, if $D^{*\,q\pm1}$ can swiftly relax to $D^{\,q\pm1}$, the situation depicted in \cref{fig:potential_transitions}c emerges. 
Here the `best of both' is obtained, with electron capture proceeding via the TL nearest the conduction band and hole capture via the TL nearest the valence band, likely expediting the recombination process as observed in Refs. \citenum{kavanagh_rapid_2021,alkauskas_role_2016,guo_nonradiative_2021,yang_non-radiative_2016}. 
Crucially, the introduction of the internal conversion to the recombination kinetics means that the timescale of the $D^{*\,q\pm1} \rightarrow D^{\,q\pm1}$ transition, relative to the slowest charge capture process, will dictate whether the metastable defect accelerates or retards the overall recombination rate.
Given typical effective vibrational (attempt) frequencies of $\sim0.5-$\SI{10}{THz},\cite{kavanagh_rapid_2021,yang_non-radiative_2016,yang_first-principles_2015} internal conversion is unlikely to be the rate-limiting step in a fast recombination cycle if the transition energy barrier $\Delta E$ is less than $0.2-$\SI{0.4}{eV} (\cref{sisubsec:ic_vs_cc_rates}$^\dag$).

Another important consideration is that for the metastable configuration to enter the recombination process, the transition level $\varepsilon(q/q^*\pm1)$ must lie within the band gap formed between the valence band maximum (VBM) and conduction band minimum (CBM):
\begin{equation}
    E_{\textrm{VBM}} < \varepsilon(q/q^*\pm1) < E_{\textrm{CBM}}
\end{equation}
Using \cref{eq:capture_into_meta_general} to substitute for $\varepsilon(q/q^*\pm1)$:
\begin{equation}
    E_{\textrm{VBM}} < \varepsilon(q/q\pm1) \mp \Delta E < E_{\textrm{CBM}}
\end{equation}
Thus for electron capture we have:
\begin{equation}
    \varepsilon(q/q-1) + \Delta E < E_{\textrm{CBM}}
\end{equation}
\begin{equation}\label{eq:e_capture_into_meta_limit}
    e^-\textrm{ capture} \Longrightarrow\ \ \Delta E < E_{\textrm{CBM}} - \varepsilon(q/q-1)
\end{equation}
and for hole capture:
\begin{equation}
    E_{\textrm{VBM}} < \varepsilon(q/q+1) - \Delta E
\end{equation}
\begin{equation}\label{eq:h_capture_into_meta_limit}
    h^+\textrm{ capture} \Longrightarrow\ \ \Delta E < \varepsilon(q/q+1) - E_{\textrm{VBM}}
\end{equation}
Hence, the energy window between the ground-state transition level $\varepsilon(q/q\pm1)$ and the corresponding band edge sets an upper limit on the relative energy of the metastable structure, if it is to yield a transition level within the band gap and potentially affect carrier recombination (\cref{eq:h_capture_into_meta_limit,eq:e_capture_into_meta_limit}).
For instance, for $D^{\,*-1}$ in \cref{fig:metastable_TLs} to introduce a transition level within the band gap, the energy of $D^{\,*-1}$ relative to $D^{\,-1}$ must be less than the energy separation of $\mathit{\varepsilon}(0/-1)$ and the CBM.

As an example, in CdTe, there are 6 intrinsic defects (two vacancies, antisites and interstitials each), which range in charge between 0 and $\pm2$ for a Fermi level within the band gap (except \kv{Te}{Cd} which ranges from -2 to +2).\cite{yang_review_2016} 
These 20 defect species yield 14 in-gap levels involving only ground-state structures. 
With our defect structure-searching technique (see Methods),\cite{mosquera-lois_search_2021} we further identify 14 \emph{metastable} structures for these charge states, generating 32 additional charge transition levels. 
Only 12 of these structures are sufficiently low energy to yield at least one transition level within the band gap, including those for \kv{Te}{i} discussed here and \kv{V}{Cd} discussed in Ref. \citenum{kavanagh_rapid_2021}, yielding 18 transition levels involving metastable structures in the CdTe band gap.
Of course, the presence of low-energy metastable defect structures in a given material will depend on several factors, including the bulk crystal symmetry, chemical bonding, and band gap, amongst others.

We note that \cref{fig:metastable_TLs} also serves as an illustrative example of the fluctuations in charge transition level positions that temperature and strong electron-phonon coupling would engender, yielding a distribution of energy levels rather than a static trap.
Such behaviour could then influence the non-radiative recombination process in a similar fashion to the impact of metastability discussed here, with non-linear temperature dependence in capture cross-sections --- an important consideration when interpreting the results of spectroscopic defect measurements.\cite{wickramaratne_defect_2018}

\subsection{Capture \emph{From} Metastable Defects}
As illustrated by \cref{fig:potential_transitions}d,e, metastable configurations can also enter the recombination cycle through an initial thermal or optical excitation, \emph{prior} to charge capture.
Again, the metastable defect $D^{*\,q}$ could have a slower or faster capture rate than the corresponding ground-state configuration $D^{\,q}$.
In the case of faster capture ($k_{\,D^* \rightarrow D^{\,q\pm1}} > k_{\,D \rightarrow D^{\,q\pm1}}$), the metastable configuration acts as an accelerating reaction intermediate to electron-hole recombination through the path $D^{\,q} \rightarrow D^{*\,q} \rightarrow D^{\,q\pm1} \rightarrow D^{\,q}$ (\cref{fig:potential_transitions}d).
This mechanism is similar to the $D^{\,q} \rightarrow D^{*\,q\pm1} \rightarrow D^{\,q\pm1} \rightarrow D^{\,q}$ route (\cref{fig:potential_transitions}c) discussed in the previous section, though with some crucial differences in the governing kinetics (now involving an internal excitation, rather than relaxation).
If accessible metastable configurations exist for both charge states, then recombination could also proceed through the pathway shown in \cref{fig:potential_transitions}e, with the charge capture transition occurring between the metastable defects $D^{*\,q} \rightarrow D^{*\,q\pm1}$, prior to de-excitation and charge capture back to $D^{\,q}$.
While certainly possible, to our knowledge a defect centre exhibiting this behaviour has not been previously reported in the literature, which may be attributed to the recency of advancements in atomistic modelling of non-radiative carrier recombination\cite{alkauskas_first-principles_2014,shi_ab_2012,sunghyun_kim_carriercapturejl_2020,turiansky_nonrad_2021,shi_comparative_2015} and our understanding of the importance of metastable structures in these processes.\cite{kavanagh_rapid_2021,yang_non-radiative_2016,alkauskas_role_2016,guo_nonradiative_2021}
Interestingly, we note that an analogous hot-electron capture mechanism was however invoked to explain the puzzling experimental observations of charge capture behaviour for negative-U DX centres in \ce{Al_xGa_{1-x}As} in the 1980s.\cite{theis_dx_1989}

\subsubsection{Tellurium Interstitial (\kv{Te}{i}) in CdTe; Structures.~~} 
\begin{figure}[h]
    \centering
    \includegraphics[width=0.9\textwidth]{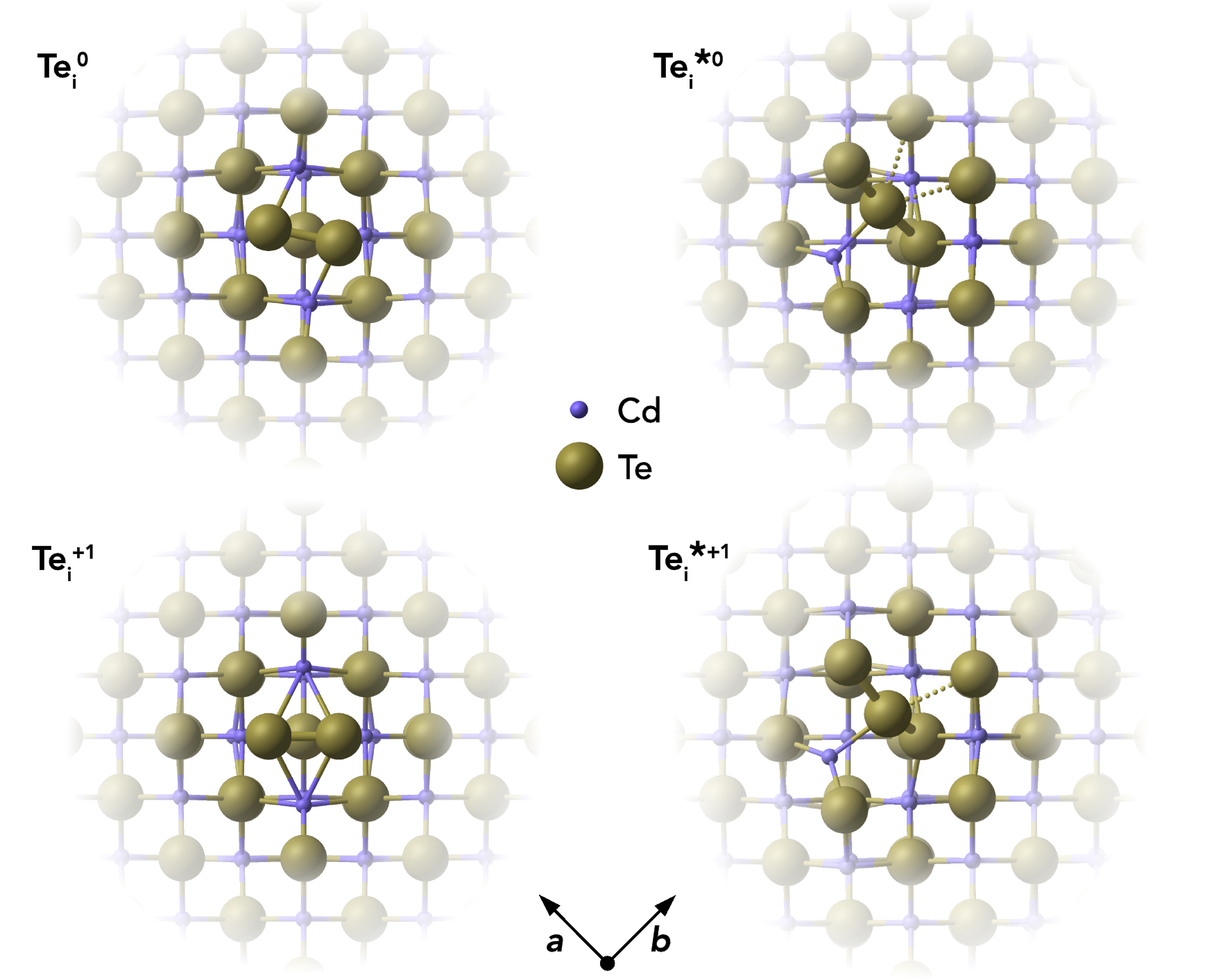}
    \caption{Atomic structures of ground-state (\kvc{Te}{i}{0/+1}) and metastable (\kvc{Te}{i}{*\,0/+1}) tellurium interstitials in CdTe, looking down the $[001]$ axis. Atoms sized according to their formal ionic radii.\cite{shannon_revised_1976} Te anions are shown in gold and Cd cations in purple.}
    \label{fig:Te_i_Strucs}
\end{figure}

\begin{figure}[h]
    \centering
    \includegraphics[width=0.75\textwidth]{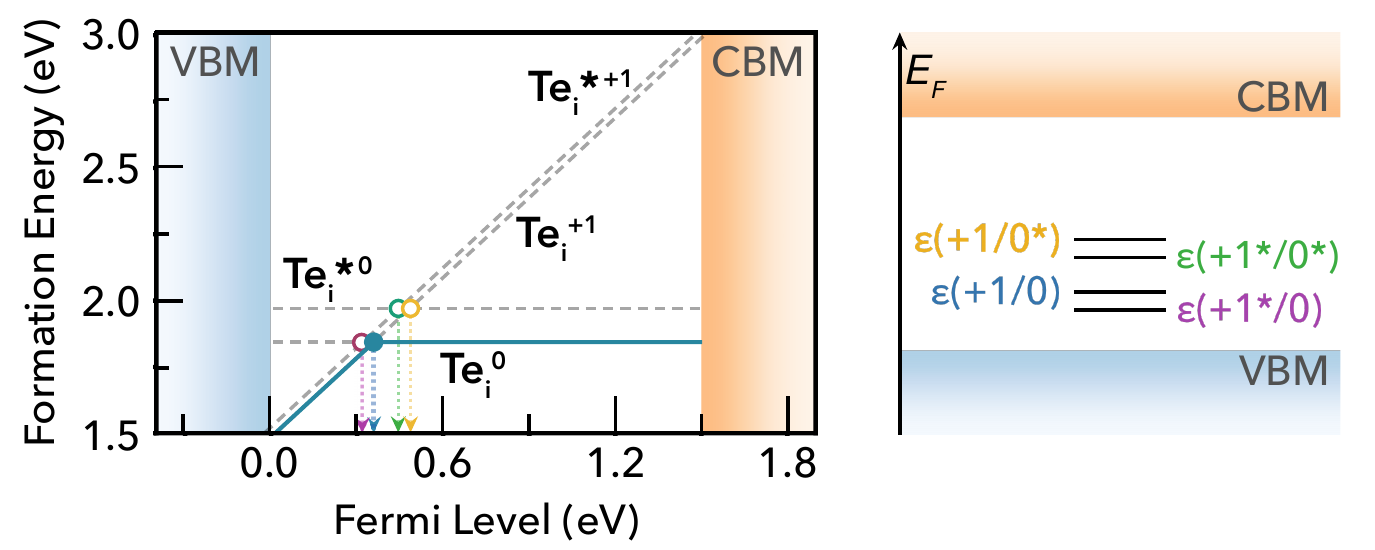}
    \caption{\textbf{(Left)} Defect formation energy diagram for the neutral and positively charged states of \kv{Te}{i} in CdTe, under Te-rich conditions ($\mu_{Te} = 0$). The equilibrium charge state is shown in solid blue, and higher energy states in dashed grey. \textbf{(Right)} Vertical energy band diagram showing the \kv{Te}{i} $(+^{(*)}/0^{(*)})$ charge transition levels in CdTe.}
    \label{fig:Te_i_TL_Diagram}
\end{figure}

In our calculations of point defects in CdTe, we found the low-energy tellurium interstitial (\kv{Te}{i}) to be an exemplar of this charge capture phenomenon.
The individual defect structures and corresponding transition level diagram for \kv{Te}{i} in CdTe are shown in \cref{fig:Te_i_Strucs} and \cref{fig:Te_i_TL_Diagram} respectively.

We find that \kv{Te}{i} has two low-energy structural arrangements for both the neutral and positively charged interstitial, resulting in four potential charge transitions (i.e. intersection points on the transition level diagram in \cref{fig:Te_i_TL_Diagram}).
In both the neutral and positively-charged ground-states, the interstitial \kv{Te}{i} displaces a host Te atom to form a split-interstitial Te-Te dimer (\cref{fig:Te_i_Strucs}), as first noted by \citet{du_native_2008}
For \kvc{Te}{i}{+1}, the dimer bond is directed along the $\langle110\rangle$ crystal direction, adopting a $C_{2v}$ point group symmetry.
For \kvc{Te}{i}{0}, the dimer bond is twisted slightly ($12\si{\degree}$) about the $[001]$ axis, thus reducing to $C_{2}$ point group symmetry with the removal of the $\{110\}$ mirror plane.

A standard \textit{ab initio} geometry relaxation from the high-symmetry tetrahedral interstitial coordination does not find the neutral ground-state \kvc{Te}{i}{0\!}, but rather converges to a higher-energy structure \kvc{Te}{i}{*\,0} with a formation energy only \SI{127}{meV} higher than the split-interstitial dimer ground-state (\cref{fig:Te_i_TL_Diagram}).
This configuration exhibits two short (\SI{2.90}{\angstrom}) and two long (\SI{3.86}{\angstrom}) \kv{Te}{i}-Te distances, and an apical \kv{Te}{i}-Cd bond (\SI{2.66}{\angstrom}) yielding $C_{2v}$ point symmetry (\cref{fig:Te_i_Strucs}).
In fact, this structure is a saddle point on the PES, representing the transition state of the \kv{Te}{i} interstitial diffusion mechanism (i.e. hopping of the split-interstitial dimer).\cite{ma_correlation_2014,selvaraj_passivation_2021,yang_first-principles_2015}
We thus determine a diffusion barrier of \SI{0.13}{eV} for \kvc{Te}{i}{0} using the HSE(34.5\%)+SOC functional (\cref{sifig:Te_i_NEBs}$^\dag$), similar to the literature value of \SI{0.09}{eV} calculated using both GGA (PBE) and hybrid (HSE06) DFT.\cite{ma_correlation_2014,yang_first-principles_2015}
A similar metastable structure exists for the positively-charged interstitial \kvc{Te}{i}{*\,+1\!}, where the partial occupancy of the defect level induces a Jahn-Teller splitting of the two long \kv{Te}{i}-Te bonds, thus reducing to $C_s$ point symmetry (\cref{fig:Te_i_Strucs}).
Here, the metastable configuration \kvc{Te}{i}{*\,+1} is only \SI{36}{meV} higher in energy than the ground-state \kvc{Te}{i}{+1\!}, now exhibiting a local stability window of \SI{40}{meV} (\cref{sifig:Te_i_NEBs}$^\dag$).

\subsubsection{\kv{Te}{i} in CdTe; Site Degeneracies.~~} Beyond the introduction of additional potential recombination pathways, the varying symmetries of the \kv{Te}{i} configurations yields differing site degeneracies for these structures.
Site degeneracy is inversely proportional to the number of symmetry operations, and so low-symmetry, distorted defect structures can possess much larger degeneracies than high-symmetry defects.
Given the linear relationship between defect concentration and configurational degeneracy $N \propto g$, these effects are not insignificant, and can alter the defect concentrations by over an order of magnitude in certain cases.
For \kvc{Te}{i}{+1\!}\nolinebreak, there are six equivalent $\langle110\rangle$ directions along which the split-interstitial dimer ($C_{2v}$ symmetry) can be oriented, which in combination with a spin degeneracy of 2 from having an unpaired electron yields a total configurational degeneracy $g_{+1} = 12$.
There are two equivalent clockwise/anti-clockwise rotations about the $[001]$ axis which give the neutral \kvc{Te}{i}{0} twisted dimer ($C_2$ symmetry)(\cref{fig:Te_i_Strucs}), and no unpaired electrons, which respectively double and half the degeneracy thus giving $g_{0} = g_{+1} = 12$.
For the saddle-point neutral defect \kvc{Te}{i}{*\,0\!}, the $C_{2v}$ structure involves two Te neighbours moving closer to the interstitial atom, and two moving away (\cref{fig:Te_i_Strucs}), giving a configurational degeneracy of $g_{0^*} = \binom{4}{2} = 6$.
The splitting of the two long Te-Te bonds in the $C_s$ \kvc{Te}{i}{*\,+1} structure (\cref{fig:Te_i_Strucs}) further doubles the number of equivalent structural configurations, yielding $g_{+1^*} = 24$ when also accounting for the spin degeneracy factor.

The consequences of degeneracy mismatches between ground and metastable defect structures can be exemplified by the temperature-dependent `reduced energy' $E_r$:\cite{grau-crespo_symmetry-adapted_2007,landsberg_recombination_1992,hayes_defects_1985}
\begin{equation}
    N_D\ =\ g N_s\,\textrm{exp}(\frac{-\Delta H}{k_B T})\ =\ N_s\,\textrm{exp}(\frac{-E_r}{k_B T})
\end{equation}
\begin{equation}
    E_r = \Delta H - k_B T\,\ln(g) 
\end{equation}
where $N_D$ is the defect concentration, $g$ is the site degeneracy, $N_s$ is the concentration of lattice sites and $\Delta H$ is the defect formation enthalpy.
The `reduced energies' for \kv{Te}{i} are given in \cref{tab:Te_i_Struc_Parameters}.
The small reduced energy difference $E_r(\SI{300}{K}) = \SI{18}{meV}$ between ground and metastable configurations for \kvc{Te}{i}{+1} corresponds to an equilibrium concentration ratio of only [\kvc{Te}{i}{*\,+1}]/[\kvc{Te}{i}{\,+1}]$ = \exp{\frac{-\Delta E_r}{k_B T}} = 0.5$ at room temperature, while for \kvc{Te}{i}{0} the greater degeneracy of the ground-state disfavours the metastable configuration.
Notably, such entropic effects become even more significant at elevated temperatures (for example during crystal growth and annealing where defects are initially introduced --- strongly affecting the resulting self-consistent Fermi level) and for cases with larger degeneracy differences (potentially over an order of magnitude).
This phenomenon is another important consequence of symmetry-breaking and metastable configurations for point defects in semiconductors, which can significantly impact bulk defect concentrations and thus Fermi level position, in addition to the effects on non-radiative recombination discussed here.

\vspace{0.3cm} 
\renewcommand{\arraystretch}{1.25}
\begin{table}[H]
\small
\caption{~Point group symmetries, site, spin and total (configurational) degeneracies ($g_{\textrm{Site}}$, $g_{\textrm{Spin}}$ and $g_{\textrm{Total}}$), relative formation enthalpies ($\Delta H$) and room temperature `reduced energies' ($E_r(T=\SI{300}{K})$) for tellurium interstitials \kvc{Te}{i}{(*)\,0/+1} in CdTe} 
\centering
\label{tab:Te_i_Struc_Parameters}
\begin{tabular}{c c c c c c c} 
\hline\hline
Defect & Point Symmetry & $g_{\textrm{Site}}$ & $g_{\textrm{Spin}}$ & $g_{\textrm{Total}}$ & $\Delta H$ & $E_r(T=\SI{300}{K})$
\\ \hline
--- & --- & --- & --- & --- & \si{meV} & \si{meV}
\\ \hline
\kvc{Te}{i}{0} & $C_2$ & 12 & 1 & 12 & 0 & 0 \\
\kvc{Te}{i}{*\,0} & $C_{2v}$ & 6 & 1 & 6 & 127 & 145 \\
\Xhline{0.5\arrayrulewidth} 
\kvc{Te}{i}{+1} & $C_{2v}$ & 6 & 2 & 12 & 0 & 0 \\
\kvc{Te}{i}{*\,+1} & $C_s$ & 12 & 2 & 24 & 36 & 18
\end{tabular}
\end{table} 


\subsubsection{\kv{Te}{i} in CdTe; Charge Capture.~~} 
\begin{figure}[h]
\centering
\includegraphics[width=\textwidth]{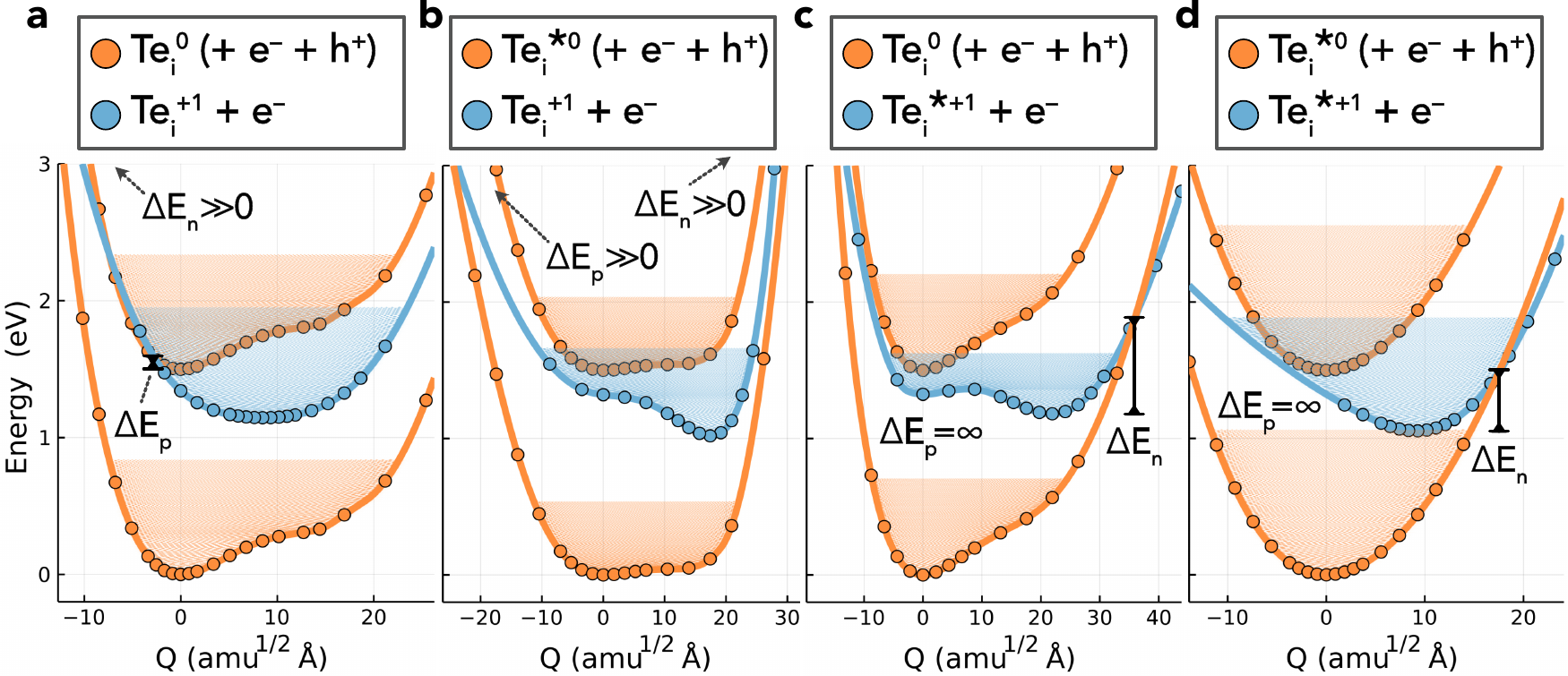}
\caption{Potential energy surfaces (PESs) corresponding to the four \kv{Te}{i} $(+^{(*)}/0^{(*)})$ charge transition levels in CdTe. Filled circles denote datapoints calculated with hybrid DFT including spin-orbit coupling, and the solid lines are spline interpolations. The low-lying effective vibrational states are shown by the shaded regions, and $\Delta E_{n/p}$ represents the classical energy barrier to the capture transition. $Q$ is the 1D structural coordinate along the path between equilibrium configurations, given in units of mass-weighted displacement. Transitions from the \emph{upper} orange to blue PESs (\kvc{Te}{i}{0} $\rightarrow$ \kvc{Te}{i}{+1}) correspond to hole capture, while those from blue to lower orange correspond to electron capture (\kvc{Te}{i}{+1} $\rightarrow$ \kvc{Te}{i}{0})} 
\label{fig:Te_i_CC_PESs}
\end{figure}

To determine the non-radiative electron/hole capture rates of the \kv{Te}{i} defect levels, we employed the one-dimensional configuration coordinate model of \citeauthor{alkauskas_first-principles_2014}\cite{alkauskas_first-principles_2014} as implemented in the \texttt{CarrierCapture.jl} package.\cite{sunghyun_kim_carriercapturejl_2020}
The potential energy surfaces (PESs) for \kv{Te}{i} along the structural paths (configuration coordinates $Q$) between neutral and positive configurations are shown in \cref{fig:Te_i_CC_PESs}.

\begin{figure}[!b]
\centering
\includegraphics[width=\textwidth]{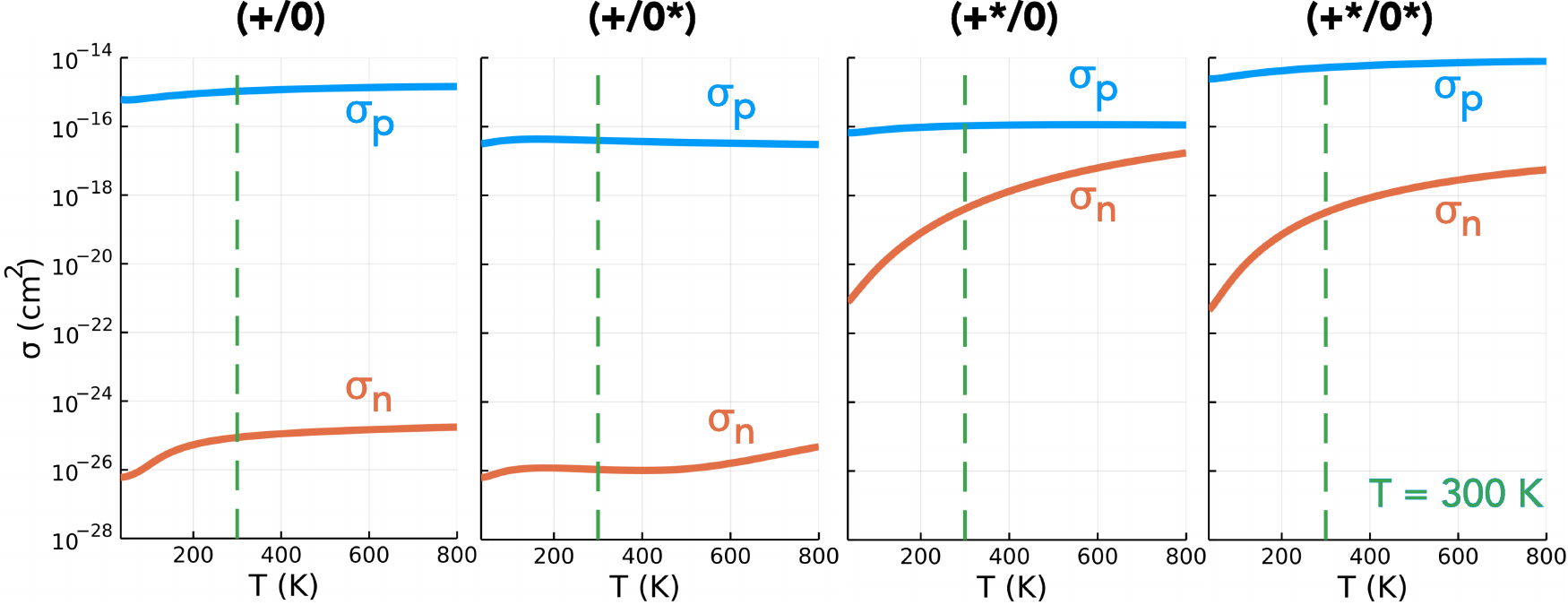}
\caption{Electron and hole capture cross-section ($\sigma_n$ and $\sigma_p$) as functions of temperature $T$ for the \kv{Te}{i} $(+^{(*)}/0^{(*)})$ charge transition levels in CdTe. Room temperature ($T =$ \SI{300}{K}) denoted by the dashed green line as a guide for typical solar cell operating conditions}
\label{fig:Te_i_Csigma}
\end{figure}

By calculating the electron-phonon coupling ($W_{if} =
\left\langle\Psi_{i}\left|\partial \hat{H} / \partial Q\right| \Psi_{f}\right\rangle$) and solving the 1D Schrödinger equation to determine the effective vibrational wavefunctions for the calculated energy surfaces, the non-radiative recombination activity of the corresponding defect transition levels (TLs) may be computed, with the results for \kv{Te}{i} provided in \cref{tab:Te_i_CC_Parameters} and \cref{fig:Te_i_Csigma}.
The relationship between the calculated PESs and resultant capture rates can be intuitively understood through the classical energy barriers to charge capture, denoted by $\Delta E_{n/p}$ in \cref{fig:Te_i_CC_PESs}.
A small capture barrier $\Delta E_{n/p}$ and/or close overlap of the defect PESs near the equilibrium position typically yields fast carrier capture, while large barriers and/or significant separation of PESs results in slow capture rates.

Inspecting the defect PESs in \cref{fig:Te_i_CC_PESs} and corresponding capture rates in \cref{tab:Te_i_CC_Parameters} \& \cref{fig:Te_i_Csigma}, we firstly note that all four potential transition pathways exhibit fast \emph{hole} capture; \kvc{Te}{i}{(*)\,0} + $h^+\rightarrow$ \kvc{Te}{i}{(*)\,+1}, with close overlap of the PESs about the upper orange \kvc{Te}{i}{(*)\,0} minimum in each case.
This behaviour is consistent with the conventional rationale of fast hole capture for defect levels located near the VBM, as is the case here (\cref{fig:Te_i_TL_Diagram}).
The large hole capture cross-sections $\sigma_p \sim 10^{-15}\,$\si{cm^2} --- of similar magnitude to its atomic cross-section $\sigma_{Te} = \pi(a_{Te})^2 \simeq \SI{1.5e-15}{cm^2}$ --- classify \kv{Te}{i} as a `giant' hole trap.\cite{pauling_nature_1961,stoneham_theory_2001,kim_upper_2020} 
On the other hand, we witness negligible \emph{electron} capture due to high barriers and low PES overlap for the $(+/0)$ and $(+/0^*)$ transition levels (\cref{fig:Te_i_CC_PESs}a,b), with tiny capture cross-sections $\sigma_n \sim 10^{-26}\,$\si{cm^2}.
It is only when \kvc{Te}{i}{*\,+1} is involved, as in the $(+^*/0)$ and $(+^*/0^*)$ levels in \cref{fig:Te_i_CC_PESs}c,d, that a tractable electron capture barrier $\Delta$E\sub{n} emerges, yielding an increase in the electron capture rate by over 6 orders of magnitude to $\sigma_n \sim 10^{-19}\,$\si{cm^2} (\cref{tab:Te_i_CC_Parameters,fig:Te_i_Csigma}).
To contextualise, we note that `giant' traps or `killer' defects are typically classified as those with capture cross-sections $\sigma \sim 10^{-15}-10^{-13}\,\si{cm^2}$, for moderate traps $\sigma \sim 10^{-20}-10^{-15}\,\si{cm^2}$ and weak traps $\sigma < 10^{-20}\,\si{cm^2}$.\cite{stoneham_theory_2001,alkauskas_first-principles_2014,kim_upper_2020}
Of course, the overall rate of capture ($R$) remains dependent on the defect and carrier concentrations ($N_D$ and $n$) in the material ($R \propto N_D \sigma_n$), meaning defects with moderate capture cross-sections can act as performance-limiting species if they form easily in the bulk, as is the case for \kv{Te}{i} in commercial Te-rich CdTe.

\begin{figure*}[t]
\begin{minipage}[t]{\linewidth}
\renewcommand{\arraystretch}{1.25}
\begin{table}[H]
\small
\caption{~Transition level position (TL), mass-weighted displacement ($\Delta Q$), classical capture barrier ($\Delta E$), capture pathway degeneracy ($g$\sub{Capture}),$^a$ electron-phonon coupling ($W_{if}$), carrier capture coefficients ($C_{n/p}$) and cross-sections $\sigma_{n/p}$ for the \kv{Te}{i} $(+^{(*)}/0^{(*)})$ defect levels in CdTe, at temperature $T\ =$ \SI{300}{K}} 
\centering
\label{tab:Te_i_CC_Parameters}
\begin{tabular}{c c c c c c c c} 
\hline\hline
TL (wrt. VBM) & $\Delta Q$ & Carrier & $\Delta E$ & $g$\sub{Capture}\super{\textit{a}} & $W_{if}$ & $C_{n/p}$ (\SI{300}{K}) & \addstackgap{$\sigma_{n/p}$ (\SI{300}{K})} 
\\ \hline
\si{eV} & \si{amu^{1/2} \angstrom} & --- & eV & --- & \si{eV / amu^{1/2} \angstrom} & \si{cm^{3}/s} & \addstackgap{\si{cm^{2}}} 
\\ \hline
\multirow{2}{*}{$(+/0)$ @ \SI{0.35}{eV}} & \multirow{2}{*}{\num{8.47}} & $e^-$ & \num{>4} & 2 & \num{2.4e-2} & \num{3.3e-18} & \num{8.7e-26} \\
& & $h^+$ & \num{0.08} & 2 & \num{1.7e-2} & \num{2.3e-8} & \num{1.0e-15} \\
\Xhline{0.5\arrayrulewidth} 
\multirow{2}{*}{$(+/0^*)$ @ \SI{0.48}{eV}} & \multirow{2}{*}{\num{17.43}} & $e^-$ & \num{2.23} & 12 & \num{2.4e-3} & \num{4.1e-19} & \num{1.1e-26} \\
& & $h^+$ & \num{3.43} & 8 & \num{1.3e-3} & \num{8.6e-10} & \num{4.0e-17} \\
\Xhline{0.5\arrayrulewidth} 
\multirow{2}{*}{$(+^*/0)$ @ \SI{0.32}{eV}} & \multirow{2}{*}{\num{21.94}} & $e^-$ & \num{0.72} & 8 & \num{2.9e-3} & \num{1.5e-11} & \num{4.1e-19} \\
& & $h^+$ & $\infty$ & 2 & \num{5.7e-3} & \num{2.3e-9} & \num{1.1e-16} \\
\Xhline{0.5\arrayrulewidth} 
\multirow{2}{*}{$(+^*/0^*)$ @ \SI{0.44}{eV}} & \multirow{2}{*}{\num{9.27}} & $e^-$ & \num{0.43} & 1 & \num{2.1e-3} & \num{1.2e-11} & \num{3.2e-19} \\
& & $h^+$ & $\infty$ & 4 & \num{2.3e-2} & \num{1.1e-7} & \num{5.2e-15}
\end{tabular}
\end{table} 
\footnotetext{$^a$ The charge capture path degeneracy is the number of equivalent paths on the PES from the initial to final state, entering as a multiplicative factor in the capture rate.}
\end{minipage} 
\end{figure*}

The slow electron capture rates of the $(+/0)$ and $(+/0^*)$ TLs rule out the possibility of significant Shockley-Read-Hall $e$-$h$ recombination through the typical two-step process (\cref{fig:potential_transitions}b), or indeed through the recently reported capture-relaxation-capture process discussed previously (\cref{fig:potential_transitions}c).
The large $\sigma_n$ of \kvc{Te}{i}{*\,+1} does, however, permit electron-hole recombination to proceed through the three-step and four-step process shown in \cref{fig:potential_transitions}d,e, involving internal excitation from the ground-state of the positive interstitial (\kvc{Te}{i}{+1} $\rightarrow$ \kvc{Te}{i}{*\,+1}), before electron capture to either the ground-state or metastable neutral interstitial (\kvc{Te}{i}{*\,+1} $+\ e^- \rightarrow$ \kvc{Te}{i}{(*)\,0}), and finally fast hole capture to complete the recombination cycle.
For the thermal excitation \kvc{Te}{i}{+1} $\rightarrow$ \kvc{Te}{i}{*\,+1\!}\nolinebreak, we calculate an energy barrier of \SI{0.08}{eV} (\cref{sifig:Te_i_NEBs}$^\dag$) and an effective vibrational attempt frequency $\nu =$ \SI{0.40}{THz}, yielding a room temperature transition rate $k_{\,+/+^*} = \SI{7.6e10}{s^{-1}}$, using transition state theory.
This internal conversion occurs far more rapidly than the rate-limiting electron capture process ($C_n^{\,+1*}\,n \simeq  (\SI{1.5e-11}{cm^3/s})(10^{12}\,\si{cm^{-3}}) \simeq 10^{-1}\,\si{s^{-1}}$), as is also the case for the barrier-less \kvc{Te}{i}{*\,0} $\rightarrow$ \kvc{Te}{i}{0} vibrational relaxation.
Further details for these calculations are provided in \cref{sisec:TST}$^\dag$.

\begin{figure}[!b]
    \noindent
    \centering
    \includegraphics[width=\textwidth]{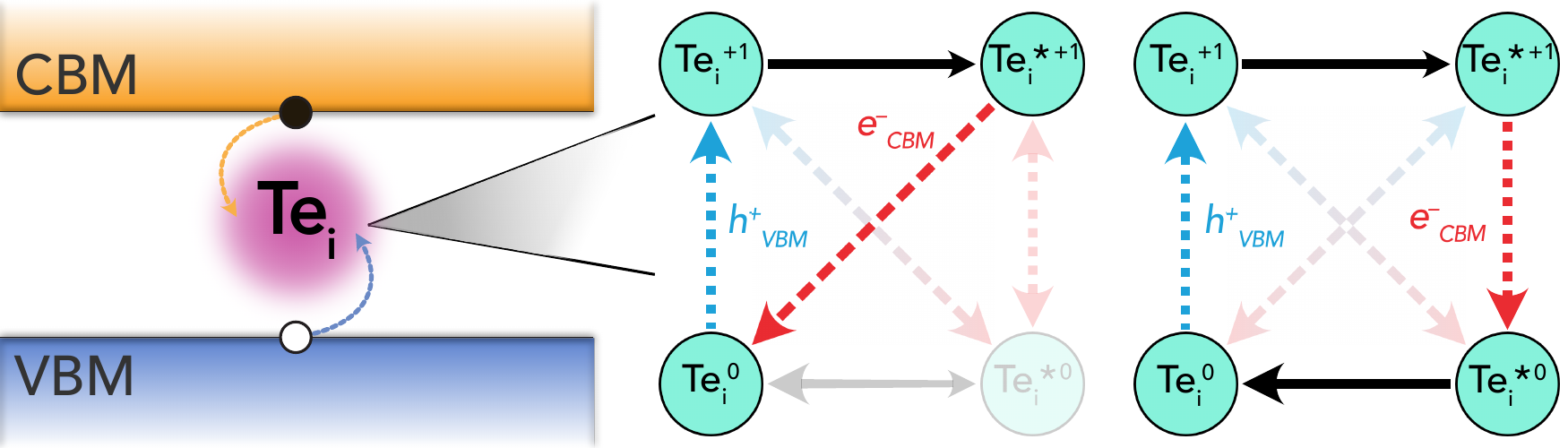}
    \caption{Schematic of the proposed non-radiative recombination mechanism at tellurium interstitials in CdTe. Red/blue arrows denote electron/hole capture, and black arrows indicate internal conversion reactions}
    \label{fig:Te_i_Recombination_Diagram}
\end{figure}

We observe facile thermal transformation between \kv{Te}{i} defect structures under typical solar cell operating conditions (black arrows in \cref{fig:potential_transitions}d,e \& \cref{fig:Te_i_Recombination_Diagram}), with electron capture by \kvc{Te}{i}{*\,+1} representing the rate-limiting step in the recombination kinetics --- now proceeding $>6$ orders of magnitude faster than for the ground-state \kvc{Te}{i}{+1}.
In summary, without the presence of the metastable \kvc{Te}{i}{*\,+1\!}, tellurium interstitials would behave as benign defect centres, only capable of capturing holes and later emitting them, reducing carrier mobilities but having no effect on recombination.
However, by introducing alternative pathways for electron capture to proceed, the low-lying metastable $+1$ structure facilitates the 3-step and 4-step recombination cycles shown in \cref{fig:potential_transitions}d,e \& \cref{fig:Te_i_Recombination_Diagram}, transforming \kv{Te}{i} into an important recombination centre in Te-rich CdTe.

\subsubsection{Capture \emph{From} Metastable Defects; General Considerations.~~} 
As before, for $\varepsilon(q^*/q\pm1)$ to enter the recombination cycle, it must be positioned within the band gap, setting an upper limit to the relative energy of the metastable structure $D^{*\,q}$ (\cref{sieq:e_capture_from_meta_limit,sieq:h_capture_from_meta_limit}$^\dag$).
The higher energy of the metastable defect ($D^{*\,q}$), now the initial state in the capture process, results in a charge transition level  $\varepsilon(q^*/q\pm1)$ which is now \emph{further} from the corresponding band edge than $\varepsilon(q/q\pm1)$ (\cref{fig:metastable_TLs,fig:Te_i_TL_Diagram,sifig:both_metastable_TLs,sieq:e_capture_from_meta,sieq:h_capture_from_meta}$^\dag$).
While a deeper defect level would typically imply slower capture velocities under the classic recombination model,\cite{huang_perovskite-inspired_2021,shockley_statistics_1952,das_what_2020,kavanagh_rapid_2021} the complexity of defect PESs results in myriad situations where this conventional wisdom no longer holds.
Indeed, we witness here that the $(+^*/0)$ charge transition level, located \emph{furthest} from the CBM of all \kv{Te}{i} $(+^{(*)}/0^{(*)})$ defect levels (\cref{fig:Te_i_TL_Diagram}),  exhibits the greatest rate of electron capture (\cref{tab:Te_i_CC_Parameters,fig:Te_i_Csigma}).
This unintuitive behaviour can be attributed to the significant structural distortion and anharmonicity along the path between these defects (\cref{fig:Te_i_CC_PESs}), highlighting the major impact that symmetry-breaking can have on the properties of defects in semiconductors.\cite{mosquera-lois_search_2021}

Accordingly, these findings yield important considerations.
Semiconductors exhibiting strong anharmonicity and mixed ionic-covalent bonding (e.g. Bi, Sb and Sn-based materials),\cite{huang_perovskite-inspired_2021,kavanagh_hidden_2021} low crystal symmetry (e.g. \ce{Sb2X3})\cite{wang_lone_2021,huang_more_2021} 
and/or compositional complexity (e.g. multinary compounds such as CZTS and double perovskites)\cite{li_effective_2019,li_bandgap_2020,kim_lone-pair_2019} are likely to manifest local minima on defect energy surfaces.
Low-energy metastable defect structures may be especially prevalent in emerging inorganic PV materials which tend to exhibit these properties, highlighting the potential importance of these species to investigations of non-radiative recombination and open-circuit voltage deficits in emerging solar cells.
Indeed, even in the case of the high-symmetry binary semiconductor CdTe, many low energy metastable structures exist, introducing 18 additional charge transition levels in the gap and accelerating the carrier trapping at \kv{Te}{i} and \kv{V}{Cd}.\cite{kavanagh_rapid_2021}
An important factor in the emergence of the thermal-excitation capture pathway here, is the high mobility of the Te interstitial in CdTe, as demonstrated by the low diffusion barriers calculated in this and previous works.\cite{ma_correlation_2014,selvaraj_passivation_2021,yang_first-principles_2015}
The facile ionic diffusion of \kv{Te}{i} is a simultaneous consequence of the soft, anharmonic PESs which cause the ready accessibility of distinct structural motifs (i.e. low-lying metastable structures), facilitating this complex recombination pathway. 
Thus, we additionally propose rapid diffusion as a potential indicator of this behaviour at other semiconductor defect centres.
Finally, we note that the very presence of metastable structures on the defect PES implies a softer and more anharmonic energy landscape, with smaller barriers between structures.
As exemplified in this and other studies,\cite{kavanagh_rapid_2021,zhang_iodine_2020,kim_anharmonic_2019,zhang_correctly_2020} such effects often yield unintuitive recombination kinetics, with fast trapping despite deep energy levels.


In conclusion, metastable defect structures impact non-radiative recombination in semiconductors.
They introduce a complex set of potential recombination paths at defects, akin to chemical reaction mechanisms, with the overall kinetics being a function of the individual transition rates. 
In addition to the thermal-relaxation pathway discussed in recent works,\cite{kavanagh_rapid_2021,yang_non-radiative_2016,alkauskas_role_2016,guo_nonradiative_2021} we demonstrate that metastable defect structures can also enter the recombination cycle via thermal excitation, provided the internal transformation energy barrier is sufficiently small.
We focus on the tellurium interstitial (\kv{Te}{i}) in CdTe solar cells as an illustrative example of this phenomenon, exhibiting complex 3-step and 4-step recombination cycles which, to our knowledge, have not been previously reported in the literature.
In addition to demonstrating the major potential impact of metastable structures on defect-mediated electron-hole recombination, \kv{Te}{i} serves a clear example of anharmonicity causing deviation from the typical trend of reduced capture rates with deeper defect levels (i.e. greater energy separation from the band edge).
Finally, we highlight implications to the broader field of photovoltaic materials research.
We pose that metastable defect structures are more important to non-radiative recombination than currently understood, particularly in the case of emerging inorganic PV materials which often exhibit reduced crystal symmetries, mixed ionic-covalent bonding, multinary compositions and/or highly-mobile defects like \kv{Te}{i}, where anharmonicity, symmetry-breaking and soft potential energy surfaces are likely to yield many low-lying, easily-accessible metastable structures, enabling the behaviour discussed here. 


\section{Methods}
Anharmonic carrier capture coefficients were calculated using  \verb|CarrierCapture.jl| ,\cite{sunghyun_kim_carriercapturejl_2020} with electron-phonon coupling matrix elements determined using the method outlined in Ref. \citenum{turiansky_nonrad_2021}.
In-depth details of the computational implementation are provided in Ref.  \citenum{kavanagh_rapid_2021}.
All of the underlying total energy calculations were performed using Density Functional Theory (DFT) within periodic boundary conditions through the Vienna \textit{Ab Initio} Simulation Package (VASP), employing Projector-Augmented Wave (PAW) pseudopotentials.\cite{kresse_ab_1993,kresse_ab_1994,kresse_efficiency_1996,blochl_projector_1994}
The screened hybrid DFT exchange-correlation functional of Heyd, Scuseria and Ernzerhof (HSE),\cite{heyd_hybrid_2003} including spin-orbit interactions (SOC), was used for all calculations.
An exact Hartree Fock exchange fraction of $\alpha_{\,\textrm{exx}}=\SI{34.5}{\%}$ was employed in the hybrid DFT model, reproducing the room temperature experimental band gap of \SI{1.5}{eV} for CdTe.
A 64-atom, \SI{13.1}{\angstrom} cubic supercell was used for defect calculations, with a well-converged \SI{450}{eV} plane-wave energy cutoff, a $2\times2\times2$ $\Gamma$-centred Monkhorst-Pack \textbf{k}-point mesh and a force convergence criterion of \SI{0.01}{eV/\angstrom} for geometry optimization.
A defect structure-searching method which aids the efficient location of ground and metastable structures for defects in solids was employed.\cite{mosquera-lois_search_2021}

\section*{Data Availability}
Data produced during this work is freely available at \url{doi.org/10.5281/zenodo.5999057}.

\section*{Acknowledgements}
Seán R. Kavanagh acknowledges the EPSRC Centre for Doctoral Training in the Advanced Characterisation of Materials (CDT-ACM)(EP/S023259/1) for funding a PhD studentship, as well as the Max-Planck-Institut für Eisenforschung and the Thomas Young Centre for funding a research visit through the `Coffee with Max Planck' and Junior Research Fellowship awards. 
DOS acknowledges support from the EPSRC (EP/N01572X/1) and from the European Research Council, ERC (Grant No. 758345).
AW acknowledges support from a National Research Foundation of Korea (NRF) grant funded by the Korean Government (MSIT) (2018R1C1B6008728). 
We acknowledge the use of the UCL Kathleen High Performance Computing Facility (Kathleen@UCL), the Imperial College Research Computing Service, and associated support services, in the completion of this work. 
Via membership of the UK's HEC Materials Chemistry Consortium, which is funded by the EPSRC (EP/L000202, EP/R029431, EP/T022213), this work used the ARCHER2 UK National Supercomputing Service (www.archer2.ac.uk) and the UK Materials and Molecular Modelling (MMM) Hub (Thomas – EP/P020194 \& Young – EP/T022213).




\footnotesize{
\bibliography{Zotero} 

\providecommand*{\mcitethebibliography}{\thebibliography}
\csname @ifundefined\endcsname{endmcitethebibliography}
{\let\endmcitethebibliography\endthebibliography}{}
\begin{mcitethebibliography}{54}
\providecommand*{\natexlab}[1]{#1}
\providecommand*{\mciteSetBstSublistMode}[1]{}
\providecommand*{\mciteSetBstMaxWidthForm}[2]{}
\providecommand*{\mciteBstWouldAddEndPuncttrue}
  {\def\EndOfBibitem{\unskip.}}
\providecommand*{\mciteBstWouldAddEndPunctfalse}
  {\let\EndOfBibitem\relax}
\providecommand*{\mciteSetBstMidEndSepPunct}[3]{}
\providecommand*{\mciteSetBstSublistLabelBeginEnd}[3]{}
\providecommand*{\EndOfBibitem}{}
\mciteSetBstSublistMode{f}
\mciteSetBstMaxWidthForm{subitem}
{(\emph{\alph{mcitesubitemcount}})}
\mciteSetBstSublistLabelBeginEnd{\mcitemaxwidthsubitemform\space}
{\relax}{\relax}

\bibitem[Huang \emph{et~al.}(2021)Huang, Kavanagh, Scanlon, Walsh, and
  Hoye]{huang_perovskite-inspired_2021}
Y.-T. Huang, S.~R. Kavanagh, D.~O. Scanlon, A.~Walsh and R.~L.~Z. Hoye,
  \emph{Nanotechnology}, 2021, \textbf{32}, 132004\relax
\mciteBstWouldAddEndPuncttrue
\mciteSetBstMidEndSepPunct{\mcitedefaultmidpunct}
{\mcitedefaultendpunct}{\mcitedefaultseppunct}\relax
\EndOfBibitem
\bibitem[Rau \emph{et~al.}(2017)Rau, Blank, M{\"u}ller, and
  Kirchartz]{rau_efficiency_2017}
U.~Rau, B.~Blank, T.~C.~M. M{\"u}ller and T.~Kirchartz, \emph{Physical Review
  Applied}, 2017, \textbf{7}, 044016\relax
\mciteBstWouldAddEndPuncttrue
\mciteSetBstMidEndSepPunct{\mcitedefaultmidpunct}
{\mcitedefaultendpunct}{\mcitedefaultseppunct}\relax
\EndOfBibitem
\bibitem[Jin \emph{et~al.}(2020)Jin, Debroye, Keshavarz, Scheblykin, Roeffaers,
  Hofkens, and Steele]{jin_its_2020}
H.~Jin, E.~Debroye, M.~Keshavarz, I.~G. Scheblykin, M.~B.~J. Roeffaers,
  J.~Hofkens and J.~A. Steele, \emph{Materials Horizons}, 2020, \textbf{7},
  397--410\relax
\mciteBstWouldAddEndPuncttrue
\mciteSetBstMidEndSepPunct{\mcitedefaultmidpunct}
{\mcitedefaultendpunct}{\mcitedefaultseppunct}\relax
\EndOfBibitem
\bibitem[Alkauskas \emph{et~al.}(2014)Alkauskas, Yan, and {Van de
  Walle}]{alkauskas_first-principles_2014}
A.~Alkauskas, Q.~Yan and C.~G. {Van de Walle}, \emph{Physical Review B}, 2014,
  \textbf{90}, 075202\relax
\mciteBstWouldAddEndPuncttrue
\mciteSetBstMidEndSepPunct{\mcitedefaultmidpunct}
{\mcitedefaultendpunct}{\mcitedefaultseppunct}\relax
\EndOfBibitem
\bibitem[Kim \emph{et~al.}(2020)Kim, Hood, {Puck van Gerwen}, Whalley, and
  Walsh]{sunghyun_kim_carriercapturejl_2020}
S.~Kim, S.~N. Hood, {Puck van Gerwen}, L.~D. Whalley and A.~Walsh,
  \emph{{{CarrierCapture}}.Jl: {{Anharmonic Carrier Capture}}}, Zenodo,
  2020\relax
\mciteBstWouldAddEndPuncttrue
\mciteSetBstMidEndSepPunct{\mcitedefaultmidpunct}
{\mcitedefaultendpunct}{\mcitedefaultseppunct}\relax
\EndOfBibitem
\bibitem[Turiansky \emph{et~al.}(2021)Turiansky, Alkauskas, Engel, Kresse,
  Wickramaratne, Shen, Dreyer, and {Van de Walle}]{turiansky_nonrad_2021}
M.~E. Turiansky, A.~Alkauskas, M.~Engel, G.~Kresse, D.~Wickramaratne, J.-X.
  Shen, C.~E. Dreyer and C.~G. {Van de Walle}, \emph{Computer Physics
  Communications}, 2021, \textbf{267}, 108056\relax
\mciteBstWouldAddEndPuncttrue
\mciteSetBstMidEndSepPunct{\mcitedefaultmidpunct}
{\mcitedefaultendpunct}{\mcitedefaultseppunct}\relax
\EndOfBibitem
\bibitem[Kim \emph{et~al.}(2019)Kim, Hood, and Walsh]{kim_anharmonic_2019}
S.~Kim, S.~N. Hood and A.~Walsh, \emph{Physical Review B}, 2019, \textbf{100},
  041202\relax
\mciteBstWouldAddEndPuncttrue
\mciteSetBstMidEndSepPunct{\mcitedefaultmidpunct}
{\mcitedefaultendpunct}{\mcitedefaultseppunct}\relax
\EndOfBibitem
\bibitem[Shi and Wang(2012)]{shi_ab_2012}
L.~Shi and L.-W. Wang, \emph{Physical Review Letters}, 2012, \textbf{109},
  245501\relax
\mciteBstWouldAddEndPuncttrue
\mciteSetBstMidEndSepPunct{\mcitedefaultmidpunct}
{\mcitedefaultendpunct}{\mcitedefaultseppunct}\relax
\EndOfBibitem
\bibitem[Kavanagh \emph{et~al.}(2021)Kavanagh, Walsh, and
  Scanlon]{kavanagh_rapid_2021}
S.~R. Kavanagh, A.~Walsh and D.~O. Scanlon, \emph{ACS Energy Letters}, 2021,
  \textbf{6}, 1392--1398\relax
\mciteBstWouldAddEndPuncttrue
\mciteSetBstMidEndSepPunct{\mcitedefaultmidpunct}
{\mcitedefaultendpunct}{\mcitedefaultseppunct}\relax
\EndOfBibitem
\bibitem[Zakutayev \emph{et~al.}(2021)Zakutayev, Major, Hao, Walsh, Tang,
  Todorov, Wong, and Saucedo]{zakutayev_emerging_2021}
A.~Zakutayev, J.~D. Major, X.~Hao, A.~Walsh, J.~Tang, T.~K. Todorov, L.~H. Wong
  and E.~Saucedo, \emph{Journal of Physics: Energy}, 2021, \textbf{3},
  032003\relax
\mciteBstWouldAddEndPuncttrue
\mciteSetBstMidEndSepPunct{\mcitedefaultmidpunct}
{\mcitedefaultendpunct}{\mcitedefaultseppunct}\relax
\EndOfBibitem
\bibitem[{Mosquera-Lois} and Kavanagh(2021)]{mosquera-lois_search_2021}
I.~{Mosquera-Lois} and S.~R. Kavanagh, \emph{Matter}, 2021, \textbf{4},
  2602--2605\relax
\mciteBstWouldAddEndPuncttrue
\mciteSetBstMidEndSepPunct{\mcitedefaultmidpunct}
{\mcitedefaultendpunct}{\mcitedefaultseppunct}\relax
\EndOfBibitem
\bibitem[Krajewska \emph{et~al.}(2021)Krajewska, Kavanagh, Zhang, Kubicki, Dey,
  Ga{\l}kowski, Grey, Stranks, Walsh, Scanlon, and
  Palgrave]{krajewska_enhanced_2021}
C.~J. Krajewska, S.~R. Kavanagh, L.~Zhang, D.~J. Kubicki, K.~Dey,
  K.~Ga{\l}kowski, C.~P. Grey, S.~D. Stranks, A.~Walsh, D.~O. Scanlon and R.~G.
  Palgrave, \emph{Chemical Science}, 2021, \textbf{12}, 14686--14699\relax
\mciteBstWouldAddEndPuncttrue
\mciteSetBstMidEndSepPunct{\mcitedefaultmidpunct}
{\mcitedefaultendpunct}{\mcitedefaultseppunct}\relax
\EndOfBibitem
\bibitem[Goes \emph{et~al.}(2018)Goes, Wimmer, {El-Sayed}, Rzepa, Jech,
  Shluger, and Grasser]{goes_identification_2018}
W.~Goes, Y.~Wimmer, A.~M. {El-Sayed}, G.~Rzepa, M.~Jech, A.~L. Shluger and
  T.~Grasser, \emph{Microelectronics Reliability}, 2018, \textbf{87},
  286--320\relax
\mciteBstWouldAddEndPuncttrue
\mciteSetBstMidEndSepPunct{\mcitedefaultmidpunct}
{\mcitedefaultendpunct}{\mcitedefaultseppunct}\relax
\EndOfBibitem
\bibitem[Lany and Zunger(2004)]{lany_metal-dimer_2004}
S.~Lany and A.~Zunger, \emph{Physical Review Letters}, 2004, \textbf{93},
  156404\relax
\mciteBstWouldAddEndPuncttrue
\mciteSetBstMidEndSepPunct{\mcitedefaultmidpunct}
{\mcitedefaultendpunct}{\mcitedefaultseppunct}\relax
\EndOfBibitem
\bibitem[Alkauskas \emph{et~al.}(2016)Alkauskas, Dreyer, Lyons, and {Van de
  Walle}]{alkauskas_role_2016}
A.~Alkauskas, C.~E. Dreyer, J.~L. Lyons and C.~G. {Van de Walle},
  \emph{Physical Review B}, 2016, \textbf{93}, 201304\relax
\mciteBstWouldAddEndPuncttrue
\mciteSetBstMidEndSepPunct{\mcitedefaultmidpunct}
{\mcitedefaultendpunct}{\mcitedefaultseppunct}\relax
\EndOfBibitem
\bibitem[Guo \emph{et~al.}(2021)Guo, Qiu, Yang, and
  Deng]{guo_nonradiative_2021}
D.~Guo, C.~Qiu, K.~Yang and H.-X. Deng, \emph{Physical Review Applied}, 2021,
  \textbf{15}, 064025\relax
\mciteBstWouldAddEndPuncttrue
\mciteSetBstMidEndSepPunct{\mcitedefaultmidpunct}
{\mcitedefaultendpunct}{\mcitedefaultseppunct}\relax
\EndOfBibitem
\bibitem[Yang \emph{et~al.}(2016)Yang, Shi, Wang, and
  Wei]{yang_non-radiative_2016}
J.-H. Yang, L.~Shi, L.-W. Wang and S.-H. Wei, \emph{Scientific Reports}, 2016,
  \textbf{6}, 21712\relax
\mciteBstWouldAddEndPuncttrue
\mciteSetBstMidEndSepPunct{\mcitedefaultmidpunct}
{\mcitedefaultendpunct}{\mcitedefaultseppunct}\relax
\EndOfBibitem
\bibitem[Das \emph{et~al.}(2020)Das, Aguilera, Rau, and
  Kirchartz]{das_what_2020}
B.~Das, I.~Aguilera, U.~Rau and T.~Kirchartz, \emph{Physical Review Materials},
  2020, \textbf{4}, 024602\relax
\mciteBstWouldAddEndPuncttrue
\mciteSetBstMidEndSepPunct{\mcitedefaultmidpunct}
{\mcitedefaultendpunct}{\mcitedefaultseppunct}\relax
\EndOfBibitem
\bibitem[Kirchartz \emph{et~al.}(2018)Kirchartz, Markvart, Rau, and
  Egger]{kirchartz_impact_2018}
T.~Kirchartz, T.~Markvart, U.~Rau and D.~A. Egger, \emph{The Journal of
  Physical Chemistry Letters}, 2018, \textbf{9}, 939--946\relax
\mciteBstWouldAddEndPuncttrue
\mciteSetBstMidEndSepPunct{\mcitedefaultmidpunct}
{\mcitedefaultendpunct}{\mcitedefaultseppunct}\relax
\EndOfBibitem
\bibitem[Zhang \emph{et~al.}(2020)Zhang, Turiansky, and {Van de
  Walle}]{zhang_correctly_2020}
X.~Zhang, M.~E. Turiansky and C.~G. {Van de Walle}, \emph{The Journal of
  Physical Chemistry C}, 2020, \textbf{124}, 6022--6027\relax
\mciteBstWouldAddEndPuncttrue
\mciteSetBstMidEndSepPunct{\mcitedefaultmidpunct}
{\mcitedefaultendpunct}{\mcitedefaultseppunct}\relax
\EndOfBibitem
\bibitem[Men{\'e}ndez-Proupin and
  Orellana(2015)]{menendezproupin_tellurium_2015-1}
E.~Men{\'e}ndez-Proupin and W.~Orellana, \emph{physica status solidi (b)},
  2015, \textbf{252}, 2649--2656\relax
\mciteBstWouldAddEndPuncttrue
\mciteSetBstMidEndSepPunct{\mcitedefaultmidpunct}
{\mcitedefaultendpunct}{\mcitedefaultseppunct}\relax
\EndOfBibitem
\bibitem[Shepidchenko \emph{et~al.}(2015)Shepidchenko, Sanyal, Klintenberg, and
  Mirbt]{shepidchenko_small_2015}
A.~Shepidchenko, B.~Sanyal, M.~Klintenberg and S.~Mirbt, \emph{Scientific
  Reports}, 2015, \textbf{5}, 1--6\relax
\mciteBstWouldAddEndPuncttrue
\mciteSetBstMidEndSepPunct{\mcitedefaultmidpunct}
{\mcitedefaultendpunct}{\mcitedefaultseppunct}\relax
\EndOfBibitem
\bibitem[Kumagai \emph{et~al.}(2021)Kumagai, Tsunoda, Takahashi, and
  Oba]{kumagai_insights_2021}
Y.~Kumagai, N.~Tsunoda, A.~Takahashi and F.~Oba, \emph{Physical Review
  Materials}, 2021, \textbf{5}, 123803\relax
\mciteBstWouldAddEndPuncttrue
\mciteSetBstMidEndSepPunct{\mcitedefaultmidpunct}
{\mcitedefaultendpunct}{\mcitedefaultseppunct}\relax
\EndOfBibitem
\bibitem[Stoneham(2001)]{stoneham_theory_2001}
A.~M. Stoneham, \emph{Theory of {{Defects}} in {{Solids}}: {{Electronic
  Structure}} of {{Defects}} in {{Insulators}} and {{Semiconductors}}}, {Oxford
  University Press}, 2001\relax
\mciteBstWouldAddEndPuncttrue
\mciteSetBstMidEndSepPunct{\mcitedefaultmidpunct}
{\mcitedefaultendpunct}{\mcitedefaultseppunct}\relax
\EndOfBibitem
\bibitem[Yang \emph{et~al.}(2015)Yang, Park, Kang, and
  Wei]{yang_first-principles_2015}
J.-H. Yang, J.-S. Park, J.~Kang and S.-H. Wei, \emph{Physical Review B}, 2015,
  \textbf{91}, 075202\relax
\mciteBstWouldAddEndPuncttrue
\mciteSetBstMidEndSepPunct{\mcitedefaultmidpunct}
{\mcitedefaultendpunct}{\mcitedefaultseppunct}\relax
\EndOfBibitem
\bibitem[Yang \emph{et~al.}(2016)Yang, Yin, Park, Ma, and
  Wei]{yang_review_2016}
J.-H. Yang, W.-J. Yin, J.-S. Park, J.~Ma and S.-H. Wei, \emph{Semiconductor
  Science and Technology}, 2016, \textbf{31}, 083002\relax
\mciteBstWouldAddEndPuncttrue
\mciteSetBstMidEndSepPunct{\mcitedefaultmidpunct}
{\mcitedefaultendpunct}{\mcitedefaultseppunct}\relax
\EndOfBibitem
\bibitem[Wickramaratne \emph{et~al.}(2018)Wickramaratne, Dreyer, Monserrat,
  Shen, Lyons, Alkauskas, and {Van de Walle}]{wickramaratne_defect_2018}
D.~Wickramaratne, C.~E. Dreyer, B.~Monserrat, J.-X. Shen, J.~L. Lyons,
  A.~Alkauskas and C.~G. {Van de Walle}, \emph{Applied Physics Letters}, 2018,
  \textbf{113}, 192106\relax
\mciteBstWouldAddEndPuncttrue
\mciteSetBstMidEndSepPunct{\mcitedefaultmidpunct}
{\mcitedefaultendpunct}{\mcitedefaultseppunct}\relax
\EndOfBibitem
\bibitem[Shi \emph{et~al.}(2015)Shi, Xu, and Wang]{shi_comparative_2015}
L.~Shi, K.~Xu and L.-W. Wang, \emph{Physical Review B}, 2015, \textbf{91},
  205315\relax
\mciteBstWouldAddEndPuncttrue
\mciteSetBstMidEndSepPunct{\mcitedefaultmidpunct}
{\mcitedefaultendpunct}{\mcitedefaultseppunct}\relax
\EndOfBibitem
\bibitem[Theis and Mooney(1989)]{theis_dx_1989}
T.~N. Theis and P.~M. Mooney, \emph{MRS Online Proceedings Library}, 1989,
  \textbf{163}, 729--740\relax
\mciteBstWouldAddEndPuncttrue
\mciteSetBstMidEndSepPunct{\mcitedefaultmidpunct}
{\mcitedefaultendpunct}{\mcitedefaultseppunct}\relax
\EndOfBibitem
\bibitem[Shannon(1976)]{shannon_revised_1976}
R.~D. Shannon, \emph{Acta Crystallographica Section A}, 1976, \textbf{32},
  751--767\relax
\mciteBstWouldAddEndPuncttrue
\mciteSetBstMidEndSepPunct{\mcitedefaultmidpunct}
{\mcitedefaultendpunct}{\mcitedefaultseppunct}\relax
\EndOfBibitem
\bibitem[Du \emph{et~al.}(2008)Du, Takenaka, and Singh]{du_native_2008}
M.-H. Du, H.~Takenaka and D.~J. Singh, \emph{Journal of Applied Physics}, 2008,
  \textbf{104}, 093521\relax
\mciteBstWouldAddEndPuncttrue
\mciteSetBstMidEndSepPunct{\mcitedefaultmidpunct}
{\mcitedefaultendpunct}{\mcitedefaultseppunct}\relax
\EndOfBibitem
\bibitem[Ma \emph{et~al.}(2014)Ma, Yang, Wei, and
  Da~Silva]{ma_correlation_2014}
J.~Ma, J.~Yang, S.-H. Wei and J.~L.~F. Da~Silva, \emph{Physical Review B},
  2014, \textbf{90}, 155208\relax
\mciteBstWouldAddEndPuncttrue
\mciteSetBstMidEndSepPunct{\mcitedefaultmidpunct}
{\mcitedefaultendpunct}{\mcitedefaultseppunct}\relax
\EndOfBibitem
\bibitem[Selvaraj \emph{et~al.}(2021)Selvaraj, Gupta, Caliste, and
  Pochet]{selvaraj_passivation_2021}
S.~C. Selvaraj, S.~Gupta, D.~Caliste and P.~Pochet, \emph{Applied Physics
  Letters}, 2021, \textbf{119}, 062105\relax
\mciteBstWouldAddEndPuncttrue
\mciteSetBstMidEndSepPunct{\mcitedefaultmidpunct}
{\mcitedefaultendpunct}{\mcitedefaultseppunct}\relax
\EndOfBibitem
\bibitem[{Grau-Crespo} \emph{et~al.}(2007){Grau-Crespo}, Hamad, Catlow, and
  de~Leeuw]{grau-crespo_symmetry-adapted_2007}
R.~{Grau-Crespo}, S.~Hamad, C.~R.~A. Catlow and N.~H. de~Leeuw, \emph{Journal
  of Physics: Condensed Matter}, 2007, \textbf{19}, 256201\relax
\mciteBstWouldAddEndPuncttrue
\mciteSetBstMidEndSepPunct{\mcitedefaultmidpunct}
{\mcitedefaultendpunct}{\mcitedefaultseppunct}\relax
\EndOfBibitem
\bibitem[Landsberg(1992)]{landsberg_recombination_1992}
P.~T. Landsberg, \emph{Recombination in {{Semiconductors}}}, 1992\relax
\mciteBstWouldAddEndPuncttrue
\mciteSetBstMidEndSepPunct{\mcitedefaultmidpunct}
{\mcitedefaultendpunct}{\mcitedefaultseppunct}\relax
\EndOfBibitem
\bibitem[Hayes and Stoneham(1985)]{hayes_defects_1985}
W.~Hayes and A.~M. Stoneham, \emph{Defects and {{Defect Processes}} in
  {{Nonmetallic Solids}}}, {Wiley}, {New York}, 1985\relax
\mciteBstWouldAddEndPuncttrue
\mciteSetBstMidEndSepPunct{\mcitedefaultmidpunct}
{\mcitedefaultendpunct}{\mcitedefaultseppunct}\relax
\EndOfBibitem
\bibitem[Pauling(1961)]{pauling_nature_1961}
L.~Pauling, \emph{The Nature of the Chemical Bond and the Structure of
  Molecules and Crystals: An Introduction to Modern Structural Chemistry},
  {Cornell University Press}, {Ithaca New York}, 3rd edn, 1961\relax
\mciteBstWouldAddEndPuncttrue
\mciteSetBstMidEndSepPunct{\mcitedefaultmidpunct}
{\mcitedefaultendpunct}{\mcitedefaultseppunct}\relax
\EndOfBibitem
\bibitem[Kim \emph{et~al.}(2020)Kim, M{\'a}rquez, Unold, and
  Walsh]{kim_upper_2020}
S.~Kim, J.~A. M{\'a}rquez, T.~Unold and A.~Walsh, \emph{Energy \& Environmental
  Science}, 2020,  1481--1491\relax
\mciteBstWouldAddEndPuncttrue
\mciteSetBstMidEndSepPunct{\mcitedefaultmidpunct}
{\mcitedefaultendpunct}{\mcitedefaultseppunct}\relax
\EndOfBibitem
\bibitem[Shockley and Read(1952)]{shockley_statistics_1952}
W.~Shockley and W.~T. Read, \emph{Physical Review}, 1952, \textbf{87},
  835--842\relax
\mciteBstWouldAddEndPuncttrue
\mciteSetBstMidEndSepPunct{\mcitedefaultmidpunct}
{\mcitedefaultendpunct}{\mcitedefaultseppunct}\relax
\EndOfBibitem
\bibitem[Kavanagh \emph{et~al.}(2021)Kavanagh, Savory, Scanlon, and
  Walsh]{kavanagh_hidden_2021}
S.~R. Kavanagh, C.~N. Savory, D.~O. Scanlon and A.~Walsh, \emph{Materials
  Horizons}, 2021, \textbf{8}, 2709--2716\relax
\mciteBstWouldAddEndPuncttrue
\mciteSetBstMidEndSepPunct{\mcitedefaultmidpunct}
{\mcitedefaultendpunct}{\mcitedefaultseppunct}\relax
\EndOfBibitem
\bibitem[Wang \emph{et~al.}(2021)Wang, Li, Kavanagh, Ganose, and
  Walsh]{wang_lone_2021}
X.~Wang, Z.~Li, S.~R. Kavanagh, A.~M. Ganose and A.~Walsh,
  \emph{arXiv:2109.08117 [cond-mat]}, 2021\relax
\mciteBstWouldAddEndPuncttrue
\mciteSetBstMidEndSepPunct{\mcitedefaultmidpunct}
{\mcitedefaultendpunct}{\mcitedefaultseppunct}\relax
\EndOfBibitem
\bibitem[Huang \emph{et~al.}(2021)Huang, Cai, Wang, Gong, Wei, and
  Chen]{huang_more_2021}
M.~Huang, Z.~Cai, S.~Wang, X.-G. Gong, S.-H. Wei and S.~Chen, \emph{Small},
  2021, \textbf{17}, 2102429\relax
\mciteBstWouldAddEndPuncttrue
\mciteSetBstMidEndSepPunct{\mcitedefaultmidpunct}
{\mcitedefaultendpunct}{\mcitedefaultseppunct}\relax
\EndOfBibitem
\bibitem[Li \emph{et~al.}(2019)Li, Yuan, Chen, Gong, and
  Wei]{li_effective_2019}
J.~Li, Z.-K. Yuan, S.~Chen, X.-G. Gong and S.-H. Wei, \emph{Chemistry of
  Materials}, 2019, \textbf{31}, 826--833\relax
\mciteBstWouldAddEndPuncttrue
\mciteSetBstMidEndSepPunct{\mcitedefaultmidpunct}
{\mcitedefaultendpunct}{\mcitedefaultseppunct}\relax
\EndOfBibitem
\bibitem[Li \emph{et~al.}(2020)Li, Kavanagh, Napari, Palgrave, {Abdi-Jalebi},
  {Andaji-Garmaroudi}, Davies, Laitinen, Julin, Isaacs, Friend, Scanlon, Walsh,
  and Hoye]{li_bandgap_2020}
Z.~Li, S.~R. Kavanagh, M.~Napari, R.~G. Palgrave, M.~{Abdi-Jalebi},
  Z.~{Andaji-Garmaroudi}, D.~W. Davies, M.~Laitinen, J.~Julin, M.~A. Isaacs,
  R.~H. Friend, D.~O. Scanlon, A.~Walsh and R.~L.~Z. Hoye, \emph{Journal of
  Materials Chemistry A}, 2020, \textbf{8}, 21780--21788\relax
\mciteBstWouldAddEndPuncttrue
\mciteSetBstMidEndSepPunct{\mcitedefaultmidpunct}
{\mcitedefaultendpunct}{\mcitedefaultseppunct}\relax
\EndOfBibitem
\bibitem[Kim \emph{et~al.}(2019)Kim, Park, Hood, and Walsh]{kim_lone-pair_2019}
S.~Kim, J.-S. Park, S.~N. Hood and A.~Walsh, \emph{Journal of Materials
  Chemistry A}, 2019, \textbf{7}, 2686--2693\relax
\mciteBstWouldAddEndPuncttrue
\mciteSetBstMidEndSepPunct{\mcitedefaultmidpunct}
{\mcitedefaultendpunct}{\mcitedefaultseppunct}\relax
\EndOfBibitem
\bibitem[Zhang \emph{et~al.}(2020)Zhang, Turiansky, Shen, and {Van de
  Walle}]{zhang_iodine_2020}
X.~Zhang, M.~E. Turiansky, J.-X. Shen and C.~G. {Van de Walle}, \emph{Physical
  Review B}, 2020, \textbf{101}, 140101\relax
\mciteBstWouldAddEndPuncttrue
\mciteSetBstMidEndSepPunct{\mcitedefaultmidpunct}
{\mcitedefaultendpunct}{\mcitedefaultseppunct}\relax
\EndOfBibitem
\bibitem[Kresse and Hafner(1993)]{kresse_ab_1993}
G.~Kresse and J.~Hafner, \emph{Physical Review B}, 1993, \textbf{47},
  558--561\relax
\mciteBstWouldAddEndPuncttrue
\mciteSetBstMidEndSepPunct{\mcitedefaultmidpunct}
{\mcitedefaultendpunct}{\mcitedefaultseppunct}\relax
\EndOfBibitem
\bibitem[Kresse and Hafner(1994)]{kresse_ab_1994}
G.~Kresse and J.~Hafner, \emph{Physical Review B}, 1994, \textbf{49},
  14251--14269\relax
\mciteBstWouldAddEndPuncttrue
\mciteSetBstMidEndSepPunct{\mcitedefaultmidpunct}
{\mcitedefaultendpunct}{\mcitedefaultseppunct}\relax
\EndOfBibitem
\bibitem[Kresse and Furthm{\"u}ller(1996)]{kresse_efficiency_1996}
G.~Kresse and J.~Furthm{\"u}ller, \emph{Computational Materials Science}, 1996,
  \textbf{6}, 15--50\relax
\mciteBstWouldAddEndPuncttrue
\mciteSetBstMidEndSepPunct{\mcitedefaultmidpunct}
{\mcitedefaultendpunct}{\mcitedefaultseppunct}\relax
\EndOfBibitem
\bibitem[Bl{\"o}chl(1994)]{blochl_projector_1994}
P.~E. Bl{\"o}chl, \emph{Physical Review B}, 1994, \textbf{50},
  17953--17979\relax
\mciteBstWouldAddEndPuncttrue
\mciteSetBstMidEndSepPunct{\mcitedefaultmidpunct}
{\mcitedefaultendpunct}{\mcitedefaultseppunct}\relax
\EndOfBibitem
\bibitem[Heyd \emph{et~al.}(2003)Heyd, Scuseria, and
  Ernzerhof]{heyd_hybrid_2003}
J.~Heyd, G.~E. Scuseria and M.~Ernzerhof, \emph{The Journal of Chemical
  Physics}, 2003, \textbf{118}, 8207--8215\relax
\mciteBstWouldAddEndPuncttrue
\mciteSetBstMidEndSepPunct{\mcitedefaultmidpunct}
{\mcitedefaultendpunct}{\mcitedefaultseppunct}\relax
\EndOfBibitem
\bibitem[Freysoldt \emph{et~al.}(2014)Freysoldt, Grabowski, Hickel, Neugebauer,
  Kresse, Janotti, and {Van de Walle}]{freysoldt_first-principles_2014}
C.~Freysoldt, B.~Grabowski, T.~Hickel, J.~Neugebauer, G.~Kresse, A.~Janotti and
  C.~G. {Van de Walle}, \emph{Reviews of Modern Physics}, 2014, \textbf{86},
  253--305\relax
\mciteBstWouldAddEndPuncttrue
\mciteSetBstMidEndSepPunct{\mcitedefaultmidpunct}
{\mcitedefaultendpunct}{\mcitedefaultseppunct}\relax
\EndOfBibitem
\bibitem[J{\'o}nsson \emph{et~al.}(1998)J{\'o}nsson, Mills, and
  Jacobsen]{jonsson_nudged_1998}
H.~J{\'o}nsson, G.~Mills and K.~W. Jacobsen, \emph{Classical and {{Quantum
  Dynamics}} in {{Condensed Phase Simulations}}}, {World Scientific}, 1998, pp.
  385--404\relax
\mciteBstWouldAddEndPuncttrue
\mciteSetBstMidEndSepPunct{\mcitedefaultmidpunct}
{\mcitedefaultendpunct}{\mcitedefaultseppunct}\relax
\EndOfBibitem
\bibitem[Vineyard(1957)]{vineyard_frequency_1957}
G.~H. Vineyard, \emph{Journal of Physics and Chemistry of Solids}, 1957,
  \textbf{3}, 121--127\relax
\mciteBstWouldAddEndPuncttrue
\mciteSetBstMidEndSepPunct{\mcitedefaultmidpunct}
{\mcitedefaultendpunct}{\mcitedefaultseppunct}\relax
\EndOfBibitem
\end{mcitethebibliography}
\bibliographystyle{rsc}
}

\pagebreak
\normalsize
\setcounter{equation}{0}
\setcounter{figure}{0}
\setcounter{table}{0}
\setcounter{section}{0}
\setcounter{secnumdepth}{3} 
\setcounter{page}{1}
\makeatletter
\renewcommand{\theequation}{S\arabic{equation}}
\renewcommand{\thefigure}{S\arabic{figure}}
\renewcommand{\thetable}{S\arabic{table}}
\renewcommand{\thesection}{S\arabic{section}}
\renewcommand{\citenumfont}[1]{S#1}

\begin{center}
\textbf{\LARGE Supplementary Material:}
\\
\LARGE{\textbf{Impact of metastable defect structures on carrier recombination in solar cells}}
\end{center}
\normalsize

\section{Charge Transition Levels Involving Metastable Defects}\label{sisec:Meta_TLs}
The formation energy $\Delta H_{D,\,q}(E_F,\mu)$ of a defect $D$ in charge state $q$, calculated using the conventional supercell method, is given by:\cite{huang_perovskite-inspired_2021,freysoldt_first-principles_2014}
\begin{equation}\label{sieq:defect_formation_energy}
\Delta H_{D,q}(E_F,\mu)\ =\ [E_{D,q}\ -\ E_H]\ -\ \sum_i n_i \mu_i\ +\ qE_F\ +\ E_{Corr}(q)
\end{equation}
where $E_{D,q}$ is the calculated energy of the defect supercell and $E_H$ is the energy of an equivalent \emph{pristine} host supercell. 
The second term ($-\ \sum_i n_i \mu_i$) accounts for the thermodynamic cost of exchanging $n_i$ atoms with their reservoir chemical potential(s) $\mu_i$, to form the defect $D_q$ from the ideal bulk material. 
Similarly, $qE_F$ represents the electron chemical potential contribution, while $E_{Corr}(q)$ is a correction for any finite-size supercell effects. 

Defect thermodynamic charge transition levels $\epsilon(q_1/q_2)$ are defined as the Fermi level position for which the formation energies of the $q_1$ and $q_2$ charge defects are equal:\cite{freysoldt_first-principles_2014}
\begin{equation}\label{eq:charge_transition_level1}
    \epsilon(q_1/q_2) = E_F;\ \ \Delta H_{D,q_1}(E_F) = \Delta H_{D,q_2}(E_F)
\end{equation}
For a Fermi level position above this defect transition level, the more negative charge state will be favoured, whereas the positive state will be favoured for a lower lying Fermi level.
Due to the charge-dependent relationship between defect formation energy and Fermi level energy ($\Delta H_{D,q}(E_F) \propto qE_F$; \cref{sieq:defect_formation_energy}), an expression for the charge transition level $\epsilon(q_1/q_2)$ can be derived following:
\begin{equation}\label{eq:charge_transition_level2}
    \Delta H_{D,\,q_1} = \Delta H_{D,\,q_1}(E_F=0) + q_1 E_F; \ \ \Delta H_{D,\,q_2} = \Delta H_{D,\,q_2}(E_F=0) + q_2 E_F
\end{equation}
Setting $\Delta H_{D,\,q_1}$ equal to $\Delta H_{D,\,q_2}$ to solve for $E_F$ (i.e. the Fermi level position for which the defect formation energies are equal $\coloneqq$ $\epsilon(q_1/q_2)$):
\begin{equation}\label{eq:charge_transition_level3}
    \Delta H_{D,\,q_1}(E_F=0) + q_1 E_F = \Delta H_{D,\,q_2}(E_F=0) + q_2 E_F
\end{equation}
\begin{equation}\label{eq:charge_transition_level4}
    \Delta H_{D,\,q_1}(E_F=0) - \Delta H_{D,\,q_2}(E_F=0) = (q_2 - q_1) E_F
\end{equation}
\begin{equation}\label{eq:charge_transition_level5}
    \longrightarrow E_F \coloneqq \epsilon(q_1/q_2) = \frac{\Delta H_{D,\,q_1}(E_F=0) - \Delta H_{D,\,q_2}(E_F=0)}{q_2 - q_1}
\end{equation}
where $\Delta H_{D,\,q}(E_F=0)$ is the formation energy of defect $D$ with charge $q$ (defined in \cref{sieq:defect_formation_energy}), when the Fermi level is at the zero-reference point (the VBM, by convention).

\begin{figure}[h] 
\centering
\includegraphics[width=0.9\textwidth]{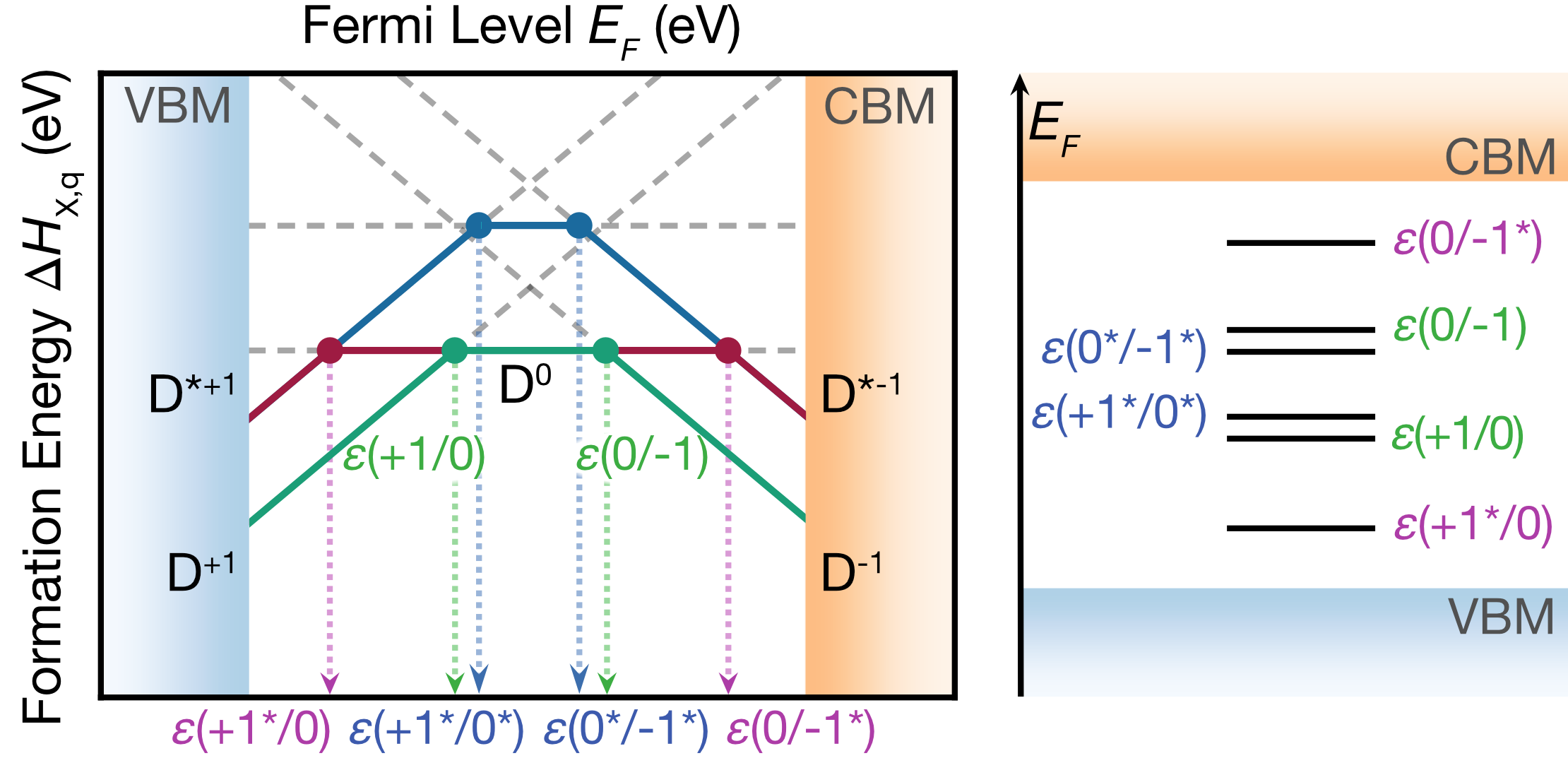}
\caption{\textbf{(Left)} Defect formation energy diagram showing the positions of charge transition levels $\varepsilon(q^{(*)}/q^{(*)}\pm1)$ involving zero (green), one (magenta) or two (blue) metastable defect structures. \textbf{(Right)} Vertical energy band diagram showing the transition level positions within the bandgap.}
\label{sifig:both_metastable_TLs}
\end{figure}

The shifts in charge transition level positions when metastable defect structures are involved are derived in \cref{sisec:Capture_into_meta,sisec:Capture_from_meta,sisec:Capture_between_meta} and illustrated in \cref{sifig:both_metastable_TLs}.

\subsection{Charge Capture \emph{Into} Metastable Defects; $D^{\,q} \rightarrow D^{\,*q\pm1}$}\label{sisec:Capture_into_meta}
Taking $q_1 = q$ and $q_2 = q\pm1$ (as charge capture is a single-carrier process) and denoting the energy of the metastable defect state $\Delta H_{D^*,\,q\pm1}$ as:
\begin{equation}
    \Delta H_{D^*,\,q\pm1} = \Delta H_{D,\,q\pm1} + \Delta E 
\end{equation}
where $\Delta E$ is the energy of the metastable defect relative to the ground-state structure, the charge transition level $\epsilon(q_1/q^*_2)$ can thus be written:
\begin{equation}
    \epsilon(q/q^*\pm1) = \frac{\Delta H_{D,\,q}(E_F=0) - \Delta H_{D^*,\,q\pm1}(E_F=0)}{q\pm1 - q} = \frac{\Delta H_{D,\,q}(E_F=0) - (\Delta H_{D,\,q\pm1}(E_F=0) + \Delta E)}{q\pm1 - q}
\end{equation}
\begin{equation}
    \epsilon(q/q^*\pm1) = \frac{\Delta H_{D,\,q}(E_F=0) - \Delta H_{D,\,q\pm1}(E_F=0)}{q\pm1 - q} - \frac{\Delta E}{q\pm1 - q}
\end{equation}
\begin{equation}
    \epsilon(q/q^*\pm1) = \epsilon(q/q\pm1) - \frac{\Delta E}{\pm1}
\end{equation}
\begin{equation}
    \epsilon(q/q^*\pm1) = \epsilon(q/q\pm1) \mp \Delta E
\end{equation}
\begin{equation}\label{sieq:e_capture_into_meta}
    e^-\textrm{ capture} \Longrightarrow\ \epsilon(q/q^*-1) = \epsilon(q/q-1) + \Delta E
\end{equation}
\begin{equation}\label{sieq:h_capture_into_meta}
    h^+\textrm{ capture} \Longrightarrow\ \epsilon(q/q^*+1) = \epsilon(q/q+1) - \Delta E
\end{equation}
Thus we witness that for charge capture \emph{into} a metastable structure, $D^{\,q} \rightarrow D^{\,*q\pm1}$, the transition level will move an energy $\Delta E$ \emph{closer to} the corresponding band-edge (i.e. to higher energy for electron capture or to lower energy for hole capture), assuming a transition level $\epsilon(q/q\pm1)$ initially located within the bandgap.

Given the requirement that $\epsilon(q/q^*\pm1)$ must lie within the bandgap in order to effectuate non-radiative recombination, we have the constraint:
\begin{equation}
    0 < \epsilon(q/q^*\pm1) < E_g
\end{equation}
\begin{equation}
    0 < \epsilon(q/q\pm1) \mp \Delta E < E_g
\end{equation}
Thus for electron capture we have:
\begin{equation}
    \epsilon(q/q-1) + \Delta E < E_g
\end{equation}
\begin{equation}
    \longrightarrow\ \Delta E < E_g - \epsilon(q/q-1)\label{sieq:e_capture_into_meta_limit}
\end{equation}
and for hole capture:
\begin{equation}
    0 < \epsilon(q/q+1) - \Delta E
\end{equation}
\begin{equation}
    \longrightarrow\ \Delta E < \epsilon(q/q+1)\label{sieq:h_capture_into_meta_limit}
\end{equation}
Therefore, we see that the energetic separation between the ground-state transition level $\epsilon(q/q\pm1)$ and the corresponding band-edge sets the energy window within which the metastable structure must lie in order to affect the electron-hole recombination process (as demonstrated in \cref{sifig:both_metastable_TLs}).

\subsection{Charge Capture \emph{From} Metastable Defects; $D^{\,*q} \rightarrow D^{\,q\pm1}$}\label{sisec:Capture_from_meta}
In the case of charge capture \emph{from} metastable structures, the same equations from the previous section hold, just with a reversal of the sign of the $\Delta E$ term:
\begin{equation}
    \epsilon(q^*/q\pm1) = \epsilon(q/q\pm1) \pm \Delta E
\end{equation}
\begin{equation}\label{sieq:e_capture_from_meta}
    e^-\textrm{ capture} \Longrightarrow\ \epsilon(q^*/q-1) = \epsilon(q/q-1) - \Delta E
\end{equation}
\begin{equation}\label{sieq:h_capture_from_meta}
    h^+\textrm{ capture} \Longrightarrow\ \epsilon(q^*/q+1) = \epsilon(q/q+1) + \Delta E
\end{equation}
Hence for charge capture \emph{from} a metastable structure, $D^{\,*q} \rightarrow D^{\,q\pm1}$, the transition level will move an energy $\Delta E$ \emph{further from} the corresponding band-edge, assuming a transition level $\epsilon(q/q\pm1)$ initially located within the bandgap.

Likewise, we obtain the following reversed constraints on the relative energy of the metastable structure, to impact carrier recombination:
\begin{equation}
    0 < \epsilon(q^*/q\pm1) < E_g
\end{equation}
\begin{equation}
    0 < \epsilon(q/q\pm1) \pm \Delta E < E_g
\end{equation}
\begin{equation}\label{sieq:e_capture_from_meta_limit}
    e^-\textrm{ capture} \Longrightarrow\ \Delta E < \epsilon(q/q-1)
\end{equation}
\begin{equation}\label{sieq:h_capture_from_meta_limit}
    h^+\textrm{ capture} \Longrightarrow\ \Delta E < E_g - \epsilon(q/q+1)
\end{equation}

\subsection{Charge Capture Between Two Metastable Defect Structures; $D^{\,*q} \longleftrightarrow D^{\,*q\pm1}$}\label{sisec:Capture_between_meta}
As demonstrated in \cref{sifig:both_metastable_TLs}, the positions of defect charge transition levels involving metastable configurations for both charge states are dictated by the balance of the relative energies of both metastable structures $D^{\,*q(\pm1)}$ with respect to their ground-state counterparts $D^{\,q(\pm1)}$, such that:
\begin{equation}
    \epsilon(q^*/q^*\pm1) = \epsilon(q/q\pm1) \pm \Delta E_{q} \mp \Delta E_{q\pm1}
\end{equation}
with the same constraint that $\epsilon(q^*/q^*\pm1)$ must lie within the bandgap in order for recombination to be energetically permitted.

%

\section{\kv{Te}{i} in CdTe: Nudged Elastic Band (NEB) Calculations}\label{sisec:NEB}
\begin{figure}[h]
\noindent
\includegraphics[width=\textwidth]{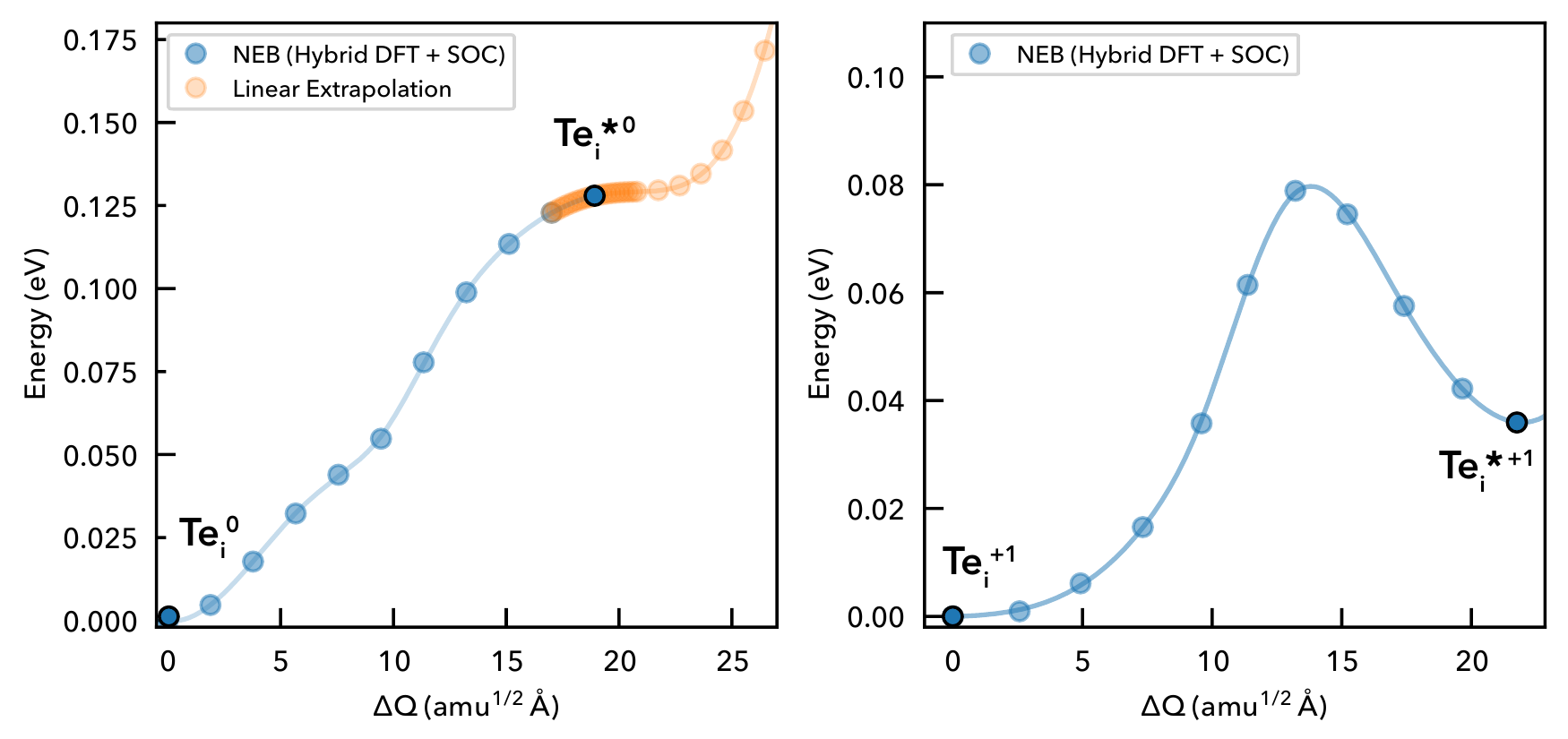}
\caption{Minimum energy paths along the potential energy surface between groundstate and metastable structures for \textbf{(left)} \kvc{Te}{i}{0} and \textbf{(right)} \kvc{Te}{i}{+1}, calculated using the Nudged Elastic Band (NEB) method.\cite{jonsson_nudged_1998} 
Filled circles represent calculated data points and the solid line is a spline fit. X-axis given in units of mass-weighted displacement.}
\label{sifig:Te_i_NEBs}
\end{figure}

\cref{sifig:Te_i_NEBs} shows the calculated minimum energy paths between the groundstate and metastable structures for \kvc{Te}{i}{0} and \kvc{Te}{i}{+1}.
As mentioned in the main text, in both cases the metastable structures represent the intermediate state in the diffusion mechanism of \kv{Te}{i} in CdTe.
Thus the energy barriers to structural transition also represent the activation energies for interstitial diffusion of tellurium in CdTe.

Using the HSE(34.5\%)+SOC hybrid DFT functional, we calculate transition energy barriers of \SI{0.13}{eV} and \SI{0.08}{eV} respectively for \kvc{Te}{i}{0} $\rightarrow$ \kvc{Te}{i}{*\,0} and \kvc{Te}{i}{+1} $\rightarrow$ \kvc{Te}{i}{*\,+1}.
We note that \citeauthor{ma_correlation_2014,yang_first-principles_2015} both found similar barriers of \SI{0.09}{eV} for the \kvc{Te}{i}{0} $\rightarrow$ \kvc{Te}{i}{*\,0} transition, with semi-local (GGA) and hybrid (HSE06) DFT respectively, and no local stability about \kvc{Te}{i}{*\,0}.
On the other hand, \citeauthor{selvaraj_passivation_2021} calculated a barrier of \SI{0.24}{eV} using local (LDA) DFT, with a \SI{0.03}{eV} window of stability for the intermediate state.

\section{\kv{Te}{i} in CdTe: Transition State Theory}\label{sisec:TST}
To estimate the rate of internal structural transformation for point defects, we can invoke Transition State Theory,\cite{vineyard_frequency_1957} which gives the rate of reaction $k$ as:
\begin{equation}\label{sieq:TST}
    k\ =\ \nu\, g\, \textrm{exp}(-\frac{\Delta E}{k_B T})
\end{equation}
where $\nu$ is the effective (vibrational) attempt frequency, $g$ is the ratio of the degeneracies of the final and initial states and $\Delta E$ is the activation energy barrier.

Using the calculated energy barrier of \SI{0.08}{eV} and effective vibrational attempt frequency $\nu =$ \SI{0.40}{THz} (from parabolic fitting of the \kvc{Te}{i}{+1} local minimum shown in \cref{sifig:Te_i_NEBs}), the \kvc{Te}{i}{+1} $\rightarrow$ \kvc{Te}{i}{*\,+1} transition is estimated to proceed at a rate:

\begin{equation*}
    k_{\,+/+^*}\ =\ (\num{4e11})\, (4)\, \textrm{exp}(-\frac{0.08}{(0.0259)(300)})\ =\ \SI{7.6e10}{s^{-1}}\ \ \textrm{@ T = }\SI{300}{K}
\end{equation*}

\subsection{Internal Conversion vs Charge Capture Rates}\label{sisubsec:ic_vs_cc_rates}
To estimate the transition barrier height at which internal conversion and charge capture would have comparable speeds, we equate their reaction rates:
\begin{equation}
    k\ =\ \nu\, g\, \textrm{exp}(-\frac{\Delta E}{k_B T})\ =\ \sigma_n\, v_{th}\, n
\end{equation}
where $\sigma_n$ is the capture cross-section, $v_{th}$ is the carrier thermal velocity and $n$ is the minority carrier concentration (minority carrier capture often represents the rate-limiting step in the electron-hole recombination cycle).
This is then rearranged to:
\begin{equation}
    \textrm{exp}(-\frac{\Delta E}{k_B T})\ =\ \frac{\sigma_n\, v_{th}\, n}{\nu\, g}
\end{equation}
\begin{equation}
    \Delta E\ =\ -k_B T \ln{\frac{\sigma_n\, v_{th}\, n}{\nu\, g}}
\end{equation}

Typical minority carrier concentrations are $n \sim \SI{1e14}{cm^{-3}}$, with velocities $v_{th} \sim \SI{1e7}{cm/s}$.
Taking these values along with the ranges $\sigma \sim 10^{-15}-10^{-13}\,\si{cm^2}$ for `killer' defect centres\cite{stoneham_theory_2001,alkauskas_first-principles_2014,kim_upper_2020} and usual attempt frequencies of $\nu \sim0.5-$\SI{10}{THz},\cite{kavanagh_rapid_2021,yang_non-radiative_2016,yang_first-principles_2015} we obtain the following upper and lower estimates for $\Delta E$:
\begin{equation}
    \Delta E_{max}\ =\ -(0.0259) \ln{\frac{(10^{-15})\, (10^{7})\, (10^{14})}{(10^{13})\, (1)}}\ =\ \SI{0.42}{eV}\ \ \textrm{@ T = }\SI{300}{K}
\end{equation}
\begin{equation}
    \Delta E_{min}\ =\ -(0.0259) \ln{\frac{(10^{-13})\, (10^{7})\, (10^{14})}{(\num{5e11})\, (1)}}\ =\ \SI{0.22}{eV}\ \ \textrm{@ T = }\SI{300}{K}
\end{equation}
Thus internal conversion is likely to be the rate-limiting step in the recombination cycle of a fast-capturing defect if the transition energy barrier $\Delta E$ is greater than \SI{0.4}{eV}, while barriers less than \SI{0.2}{eV} should yield structural conversion rates far quicker than charge capture.
For barriers between these extremal limits, which process becomes the rate-limiting step will depend on the specific parameters for that system.

\end{document}